\documentclass[twocolumn]{aastex62}

\usepackage{graphicx,epsfig,natbib,color}
\usepackage{amsmath,amssymb,bm}
\usepackage{graphicx}
\usepackage{epsfig}
\usepackage{natbib}

\definecolor{DarkGreen}{rgb}{0.0,0.45,0.0}     
\definecolor{ForrestGreen}{rgb}{0.133,0.545,0.133}
\definecolor{DarkMagenta}{rgb}{0.45,0.0,0.45}
\definecolor{DarkOrange}{rgb}{0.6,0.4,0}

\shorttitle{Initiation and Early Kinematic Evolution}
\shortauthors{Cheng et al.}
\begin{document}

\title{\textbf{Initiation and Early Kinematic Evolution of Solar Eruptions}}

\correspondingauthor{Xin Cheng}\email{xincheng@nju.edu.cn}

\author[0000-0003-2837-7136]{X. Cheng}
\affil{School of Astronomy and Space Science, Nanjing University, Nanjing 210023, China\\}
\affil{Max Planck Institute for Solar System Research, Gottingen, 37077, Germany\\}
\affil{Key Laboratory of Modern Astronomy and Astrophysics (Nanjing University), Ministry of Education, Nanjing 210093, China\\}

\author{J. Zhang}
\affil{Department of Physics and Astronomy, George Mason University, Fairfax, VA 22030, USA\\}

\author[0000-0002-5740-8803]{B. Kliem}
\affil{Institute of Physics and Astronomy, University of Potsdam, D-14476 Potsdam, Germany\\}

\author[0000-0003-3843-3242]{T. {T{\"o}r{\"o}k}}
\affil{Predictive Science Inc., 9990 Mesa Rim Rd., Ste. 170, San Diego, CA 92121, USA\\}

\author{C. Xing}
\affil{School of Astronomy and Space Science, Nanjing University, Nanjing 210023, China\\}
\affil{Key Laboratory of Modern Astronomy and Astrophysics (Nanjing University), Ministry of Education, Nanjing 210093, China\\}

\author{Z. J. Zhou}
\affil{School of Atmospheric Sciences, Sun Yat-sen University, Zhuhai, Guangdong, 519000, China\\}

\author{B. Inhester}
\affil{Max Planck Institute for Solar System Research, Gottingen, 37077, Germany\\}

\author[0000-0002-4978-4972]{M. D. Ding}
\affil{School of Astronomy and Space Science, Nanjing University, Nanjing 210023, China\\}
\affil{Key Laboratory of Modern Astronomy and Astrophysics (Nanjing University), Ministry of Education, Nanjing 210093, China\\}

\begin{abstract}

We investigate the initiation and early evolution of 12 solar eruptions, including six active region hot channel and six quiescent filament eruptions, which were well observed by the \textsl{Solar Dynamics Observatory}, as well as by the \textsl{Solar TErrestrial RElations Observatory} for the latter. The sample includes one failed eruption and 11 coronal mass ejections, with velocities ranging from 493 to 2140~km~s$^{-1}$. A detailed analysis of the eruption kinematics yields the following main results.
(1) The early evolution of all events consists of a slow-rise phase followed by a main-acceleration phase, the height-time profiles of which differ markedly and can be best fit, respectively, by a linear and an exponential function. This indicates that different physical processes dominate in these phases, which is at variance with models that involve a single process.
(2) The kinematic evolution of the eruptions tends to be synchronized with the flare light curve in both phases. The synchronization is often but not always close. A delayed onset of the impulsive flare phase is found in the majority of the filament eruptions (5 out of 6). This delay, and its trend to be larger for slower eruptions, favor ideal MHD instability models.
(3) The average decay index at the onset heights of the main acceleration is close to the threshold of the torus instability for both groups of events (although based on a tentative coronal field model for the hot channels), suggesting that this instability initiates and possibly drives the main acceleration. 

\end{abstract}


\section{Introduction}\label{s:intro}

Coronal mass ejections (CMEs) are the largest explosive phenomena in the solar system. Occurring in the solar atmosphere, they can eject a large quantity of plasma and magnetic flux into the interplanetary space. When the magnetized plasma arrives at the Earth, it will interact with the magnetosphere, potentially producing severe space weather effects, thus affecting the safety of human high-tech activities, especially in the outer space \citep{gosling93,webb94}. 

White-light coronagraph observations revealed that CMEs often have a three-part structure: a bright front followed by a bright core embedded in a dark cavity \citep{illing83}. The bright front originates from plasma pile-up at the front of the expanding CME \citep[e.g.,][]{vourlidas03,cheng14_tracking}. The cavity, or its central part, is usually interpreted to be a coherent helical flux rope \citep[e.g.,][]{dere99,gibson06_jgr,riley08,song17}. The bright core represents dense plasma usually attributed to an erupting filament/prominence, which is suspended in magnetic dips of a flux rope or in a sheared arcade prior to the eruption \citep[e.g.,][]{guo10_filament,inoue13,jiang14_filament,yanxl15,suyingna15}, but can also consist of swept-up loops \citep{Veronig&al2018}. Sometimes a fourth component, a shock, appears at the front and flanks of the CME, if its expansion velocity exceeds the local Alfv\'en speed \citep{vourlidas03,kwon14}.
 
The kinematic evolution of CMEs is usually comprised of three phases: a slow-rise phase of approximately uniform velocity, an impulsive main-acceleration phase, and a propagation phase with only slowly varying velocity \citep{zhang01,zhang04}. The slow-rise and main-acceleration phases are often also displayed by an associated filament/prominence eruption \cite[e.g.,][]{sterling07,sterling11}.

To understand the initiation and early evolution of CMEs, the relationship between their kinematic evolution and the light curve of their associated flares has been studied extensively. The three CME evolution phases were found to correspond to, respectively, the pre-flare phase, rise phase, and decay phase of the associated flare in soft X-rays (SXRs) \citep[][]{zhang01,zhang04,neupert01,cheng10_buildup,bein12}. This is further supported by a statistical study of a sample of 22 CMEs performed by \citet{maricic07}. However, those authors pointed out that the onset of the flare rise phase was delayed with respect to the onset of the CME main-acceleration phase in some of their events. In addition, \citet{qiu04} and \citet{temmer08,temmer10} uncovered that the acceleration of CMEs and the hard X-ray flux of the associated flares are often synchronized as well. These results strongly suggest that CMEs and flares are two distinct manifestations of the same process (or processes), which is a violent disruption of the coronal magnetic field \citep[e.g.,][]{forbes00a}.
 
The kinematic evolution of a CME low in the corona can be obtained by following the eruption of features of the pre-eruptive configuration. The most common tracers of pre-eruptive configurations are filaments/prominences, which are cool and dense plasma embedded in the hot and tenuous corona \citep{mackay10}. Statistically, over 70\% of CMEs are associated with erupting filaments/prominences \citep[e.g.,][]{munro79,webb87,gopal_2003}. In addition to the association with erupting filaments/prominences, \citet{cheng11_fluxrope} and \citet{zhang12} discovered that the pre-eruptive configuration can also manifest as a hot plasma channel (or hot blob when viewed along its axis) in the Atmospheric Imaging Assembly \citep[AIA;][]{lemen12} 131 {\AA} and 94 {\AA}  passbands. Interestingly, hot channels keep a coherent structure throughout the eruption \citep{zhang12,cheng13_driver,patsourakos13,nindos15}. This ensures that their heights (e.g., their distance from the solar surface) can be measured continuously and reliably in the whole AIA field of view.

When studying CME kinematics, the CME bright front is usually used to infer the CME height. This is, however, inappropriate when studying the early evolution of CMEs. Through analyzing a limb CME event, \citet{patsourakos10} found that the CME originated from the fast expansion of a plasma bubble \citep[also see][]{patsourakos10_genesis,wanlf16}. They showed that the kinematic evolution of the CME actually included two components, one being associated with the lifting of the CME centroid (the geometric center of the bubble), and the other with the expansion of the CME bubble. Furthermore, \citet{cheng13_driver} investigated the formation of two CMEs from erupting hot channels and found that the expansion of the channels coincided in time with the expansion of the CME bubbles. Moreover, they found that the hot channel rose faster than the front of the CME bubble during the main-acceleration phase \citep[also see][]{Veronig&al2018}. These results suggest that hot channels behave as a central engine that drives the formation and acceleration of CMEs during their early stage, and are therefore a better tracer of the CME kinematics in this phase.

Several mechanisms have been suggested to explain the initiation of CMEs \citep[e.g.,][]{green18}. One category of mechanisms includes tether-cutting reconnection \citep{moore01} and breakout reconnection \citep{antiochos99,karpen12}. The former takes place in the center of sigmoids low in the corona and transforms sheared arcades into a flux rope, which then lifts off, driven by the rope's magnetic pressure, as a consequence of reduced line tying \citep[e.g.,][]{liur10,chenhd14}. The breakout mechanism resorts to reconnection at a high-lying X-line (or null point) located between central sheared flux and overlying flux connecting the outer polarities in a quadrupolar magnetic configuration. By removing the constraint of the overlying flux, the downward tension is reduced, allowing the central flux to escape \citep[e.g.,][]{gary04,shenyuandeng12}.

A second category invokes ideal MHD instabilities such as the torus instability \citep{kliem06,olmedo10} and the helical kink instability \citep{Sakurai1976,fan03,kliem04}. The torus instability refers to the expansion instability of a toroidal current channel (flux rope), which commences if the decay index of the background field exceeds the critical value of $\sim$1.5 \citep{kliem06}. Considering a more realistic flux rope structure that resembles a line-tied partial torus, \citet{olmedo10} pointed out that the critical value depends on the ratio of the arc length of the partial torus and the circumference of a circular torus with equal radius. \citet{demoulin10} and \citet{kliem14_torus} demonstrated that the torus instability is equivalent to a catastrophic loss of equilibrium in the MHD framework \citep{forbes91,linjun02}. In addition to the torus instability, the helical kink instability can also initiate the eruption of a flux rope, if its twist number exceeds a certain threshold \citep{torok04,fan04}. The latter varies for different flux rope configurations \citep{baty01}. Once the helical kink instability takes place, the flux rope axis will writhe and present a distinct inverse-$\gamma$ or $\Omega$ morphology \citep[e.g.,][]{ji03,williams05,torok10,hassanin16,song18}. However, the observations indicate that this instability is unlikely a universal onset mechanism for solar eruptions, because a critical twist appears to be reached only in a minority of their source regions. Moreover, this instability is not generally suited as a mechanism for the main CME acceleration over a large height range, because it tends to saturate quickly.

The height-time profile of CMEs, $h(t)$, in the early phase of an eruption, and its association with the SXR flare light curve, or flux temporal profile, $F_\mathrm{SXR}(t)$, can help to differentiate between initiation models in three ways that will be addressed in the present paper. First, the existence of a break between the slow-rise and main-acceleration phases argues against the suggestion that a single process (e.g., ``runaway reconnection'') is the primary driver of the whole eruption. Second, any temporal offset between $h(t)$ and $F_\mathrm{SXR}(t)$ favors ideal MHD models if $h(t)$ is preceding, while it favors reconnection models if $F_\mathrm{SXR}(t)$ is preceding and the relevant flare onset is not masked by precursor activity. Third, any correlation between the onset of eruptions, either of the slow-rise or of the main-acceleration phase, and the threshold of the torus or helical kink instability favors the ideal MHD model, because the threshold should not play any role in the reconnection models.

There have been many investigations of CME height-time curves. Most have found an amplifying acceleration in the main-acceleration phase in basic agreement with all models, i.e., with instability, both ideal and resistive, and with the idea of runaway reconnection. Acceleration profiles close to an exponential \citep{vrsnak01,gallagher03,williams05} or close to a power law \citep{kahler88,vrsnak01,alexander02,schrijver08_filament} were typically found. \citet{schrijver08_filament} demonstrated that both functional forms can result from an ideal instability, with the exponential and power law indicating small and sizable perturbations of an initial equilibrium, respectively. This implies that the specific form of the main acceleration has little bearing on the debate about eruption models, but rather on the magnitude of the perturbation that triggers the onset of an eruption. Some authors obtained satisfactory fits to prominence height-time data assuming uniform acceleration \cite[e.g.,][]{gilbert00,gopals03}. However, this assumption implies a discontinuity in the acceleration at the onset of the main-acceleration phase, which is unphysical, so only allows one to characterize this phase roughly and in a global sense. The slow rise is often found to be nearly linear \cite[e.g.,][]{sterling07,sterling11,schrijver08_filament}. Consequently, a break in characteristic behavior between the slow-rise and main-acceleration phases is indicated by the majority of the previous studies.

\begin{figure*}
 \center {\includegraphics[width=14cm]{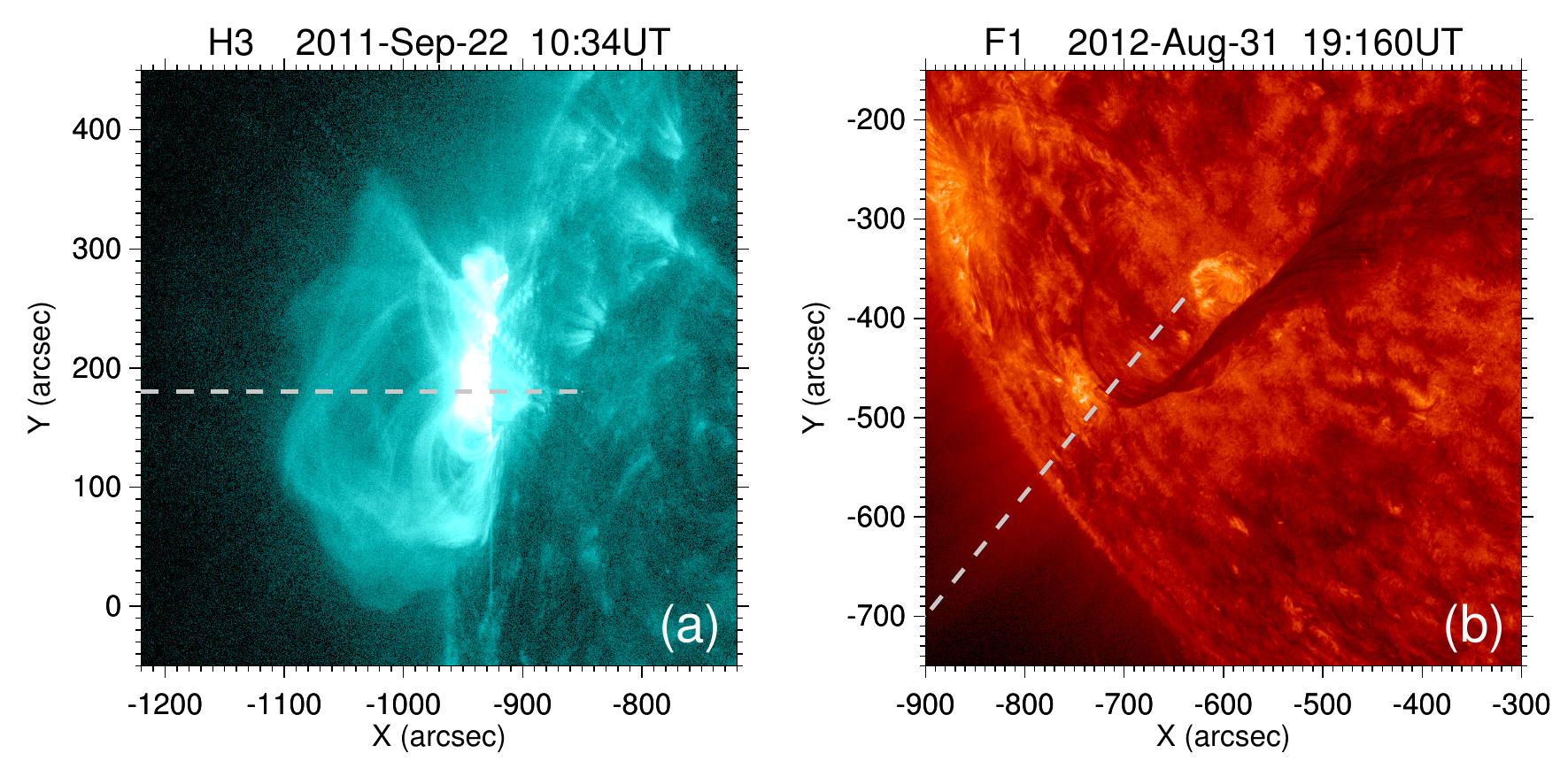}}
\caption{\textsl{SDO}/AIA 131~{\AA} and 304~{\AA} images showing (a) the erupting hot channel H3 and (b) the quiescent filament F1. The dashed lines indicate the direction of eruption.}  
\label{aia}
\end{figure*}

\begin{table*}
\caption{CME/flare properties of 12 eruption events.}
\label{tb1}{
\hspace{-0.1\textwidth}
\scalebox{0.8}[0.8]{
\begin{tabular}{cccccccc}
\\ \tableline \tableline
 Events$^{a}$   & Date  & Flare$^{b}$ & Magnitude & CME$^{c}$ &Speed$^{d}$ &Location & References$^{e}$              \\
                         &           &                    &                  &                    & [km s$^{-1}$] &              &    \\
\tableline
H1     & 2011-03-08  &Y & M1.5           & S  & 732    &S20E75 & \citet{cheng12_dem,cheng13_driver,zhang12}     \\
H2     & 2011-09-12  &Y & C9.9           & C  & -- &N20E85            & \citet{tripathi13,cheng14_kink} \\
H3     & 2011-09-22  &Y &X1.4            & S  & 1905  &N12E85 & \citet{nindos15}  \\
H4     & 2012-01-23  &Y & C?             & S  & 684     &N25W35 &  \citet{cheng13_double} \\
H5     & 2013-05-22  &Y & M5.0          & S  & 1466   &N13W80 &  \citet{liting13_homologous,cheng14_tracking} \\
H6     & 2014-02-25  &Y & X4.9           & S  & 2142  &S14E85  & \citet{chenhd14,
seaton17} \\
\tableline
F1     & 2012-08-31  &Y & C8.0           & S & 1442    &S25E40  & \citet{wood16,sinha19}  \\
F2     & 2012-11-23  &N & --                & S & 519      &S40E15  & \citet{sinha19} \\
F3     & 2013-03-16  &N & --                & S & 786      &N30W60 & -- \\
F4     & 2013-08-20  &N & --                & S & 784      &S40W00 & \citet{liting15,sinha19} \\
F5     & 2013-09-29  &Y & C1.5           & S & 1179    &N15W25 & \citet{liting15,yanxl15_quiescent_filament,palacios15,sinha19} \\
F6     & 2014-09-02  &N & --                & S & 493      &N25W10 & \citet{ouyang15} \\
\tableline
\end{tabular}}}\\

\vspace{0.01\textwidth}
Notes:\\
$^{a}$ H (F) refers to hot channel (quiescent filament) eruptions.\\
$^{b}$ Y (N) denotes a detectable (not detectable) flare in the \textsl{GOES} 1--8~{\AA} flux.\\
$^{c}$ S shows successful eruptions that produce CMEs, C denotes a failed eruption.\\
$^{d}$ The average CME speed in the LASCO field of view obtained from https://cdaw.gsfc.nasa.gov.\\
$^{e}$ Previous investigations of the event.\\
\end{table*}

Quantitative investigations of whether the slow-rise and main-acceleration phases show the same or different functional forms were, to our knowledge, presented only by \citet{kahler88}, who suggested a common power law for both, and \citet{schrijver08_filament}, who suggested different functions. Here, we first address the different findings by \citet{kahler88} and \citet{schrijver08_filament} by studying the kinematics of a larger sample of 12 events observed with high resolution and at high cadence \cite[similar to those in][]{schrijver08_filament}. To account for the broad range of CME speeds, and to permit disclosing potential differences between slow and fast CMEs, we have carefully chosen six eruptions from active regions and six from the quiet Sun. The best fits of the $h(t)$ data yield a relatively precise timing of the kinematic evolution, which is then compared with the flare light curve for each event. This might either yield a discrimination between the ideal MHD and resistive eruption models, or provide information on how early and closely the feedback between flux rope instability and reconnection is established, thereby adding to the substantial existing knowledge, which has not yet established a definite picture \citep{zhang06,maricic07,bein12}. Similarly, a relatively precise height of the onset of both the slow-rise and main-acceleration phases is obtained, which we utilize to determine whether one of these onsets is related to the threshold of the torus instability. 

We introduce the instruments in Section~\ref{s:instruments}. The criteria for choosing the sample are explained in Section~\ref{s:events}, which is followed by the results in Section~\ref{s:results}. The methods and results are discussed in Section~\ref{s:discussion}, and the conclusions are summarized in Section~\ref{s:conclusions}. 

\section{Instruments}\label{s:instruments}

The data sets are mainly from the AIA on board the \textsl{Solar Dynamics Observatory} \citep[\textsl{SDO};][]{pesnell12} which images the corona almost simultaneously at temperatures from 0.06 MK to 20 MK in ten different passbands. The temporal cadence and spatial resolution are 12 s and 1.2{\arcsec}, respectively. The two AIA high temperature passbands, 131 {\AA} and 94 {\AA}, with peak responses at temperatures of $\sim$11 MK and $\sim$7 MK, respectively, are used for identifying hot channels and tracking their evolution in the low corona; the AIA 304 {\AA} passband is for analyzing quiescent filaments. In order to determine the height of filaments in 3D, the EUVI 304 {\AA} images of the Sun Earth Connection Coronal and Heliospheric Investigation \citep[SECCHI;][]{howard08} on board the \textsl{Solar TErrestrial RElations Observatory (STEREO)} are also utilized although with a low cadence (10 min) and resolution (2.4\arcsec). The 720 s line-of-sight magnetograms of the full disk and daily updated synoptic maps provided by the Helioseismic and Magnetic Imager \citep[HMI;][]{schou12}, also on board \textsl{SDO}, are taken as the bottom boundary condition for computing a 3D coronal magnetic field model by extrapolation. We also use the Large Angle and Spectrometric Coronagraph \citep[LASCO;][]{brueckner95} on board the \textsl{Solar and Heliospheric Observatory (SOHO)} to inspect the properties of CMEs. The 1--8~{\AA} SXR flux of associated flares is provided by the \textsl{Geostationary Operational Environmental Satellite (GOES)}. 

\section{Event Selection}\label{s:events}

In this study, we collect 12 eruptive events including 6 hot channel eruptions and 6 quiescent filament eruptions. Hot channels originate from active regions with strong magnetic fields and are prone to produce fast CMEs. The visibility of the hot channels only in the AIA 131~{\AA} and 94~{\AA} passbands but not in other cooler passbands proves their high-temperature nature. Quiescent filaments are from long-time decayed active regions with weak magnetic fields and usually give rise to slow CMEs. 

We select hot channels that are mostly located at or near the solar limb. The low level of background and foreground emission ensures that the edge of the hot channels is sharp enough to allow tracking their height reliably. The high cadence of the AIA data (compared to EUVI data) yields a large number of data points, even though the hot channels evolve rapidly. This results in the most accurate height-time data currently available, which turns out to be crucial for the reliability of the fits and the derived break points and onset heights. On the other hand, reliable magnetograms can then only be obtained 3--4 days before or after the eruptions, which affects the estimates of the decay index of the coronal field at eruption onset. Therefore, we selected only events that occurred at least several days after the emergence phase of the corresponding active regions, at which time the photospheric magnetic field evolved gradually.

The quiescent filaments are selected using the catalog compiled by \citet{mcCauley15} from a longitude range of $\pm60^\circ$. This permits for daily updates of the eruption source region in the synoptic magnetograms which are used for the computation of the potential coronal field model. We require that the filaments have clear moving fronts during the eruption process, so that their height can be measured continuously and reliably. To allow determining the true height, we only select filaments with simultaneous observations of the AIA and the EUVI from two perspectives, at least during part of the rise. 

Table~\ref{tb1} shows the basic properties of the 12 events. One can see that all hot channel eruptions have an associated SXR flare and a corresponding CME, except the H2 event. \citet{cheng14_kink} and \citet{tripathi13} analyzed the H2 event in detail and found that its eruption was eventually confined by the overlying field in the high corona and thus did not produce a CME. However, it still experienced a slow-rise and then main-acceleration process during the beginning of the eruption. From Table~\ref{tb1}, it is also found that the CMEs from the hot channel eruptions do have a relatively high velocity in the range of $\approx$700--2100~km~s$^{-1}$. For the CMEs from the quiescent filaments, the velocity is inclined to be smaller, with the range of $\approx$500--1400~km~s$^{-1}$. Most of them lack detectable \textit{GOES} SXR 1--8~{\AA} flux, except the F1 and F5 eruptions, which produce two relatively fast CMEs with velocities above 1100~km~s$^{-1}$. Overall, the velocities of CMEs constituting our sample cover a large velocity range of CMEs \citep[e.g.,][]{zhang06,yashiro06}. In this sense, the results from our small sample are applicable to a broad range of events. Note that the velocity here refers to an average projected value in the LASCO field of view.

\begin{figure*}
 \center {\includegraphics[width=18cm]{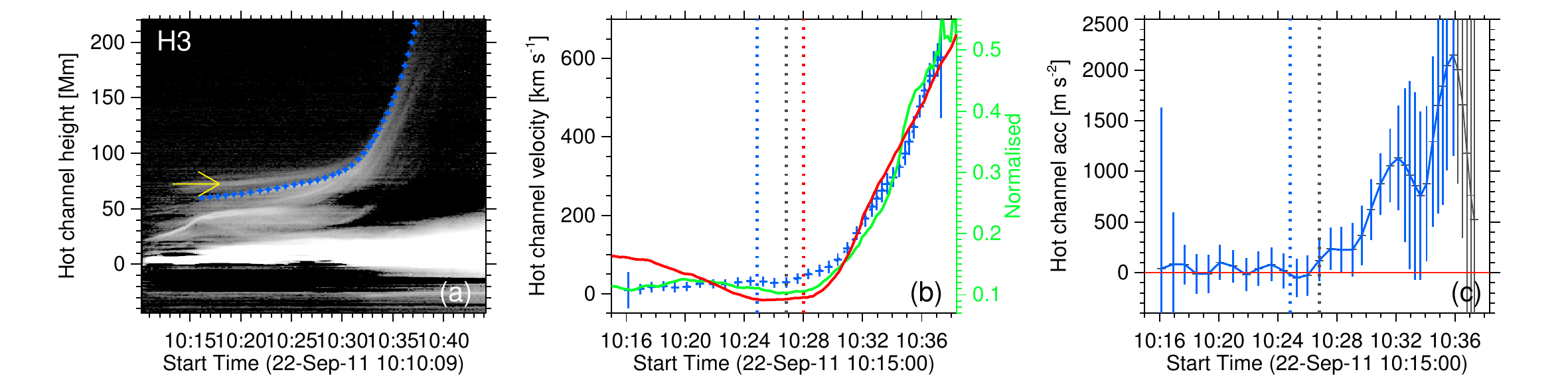}}
 \center {\includegraphics[width=18cm]{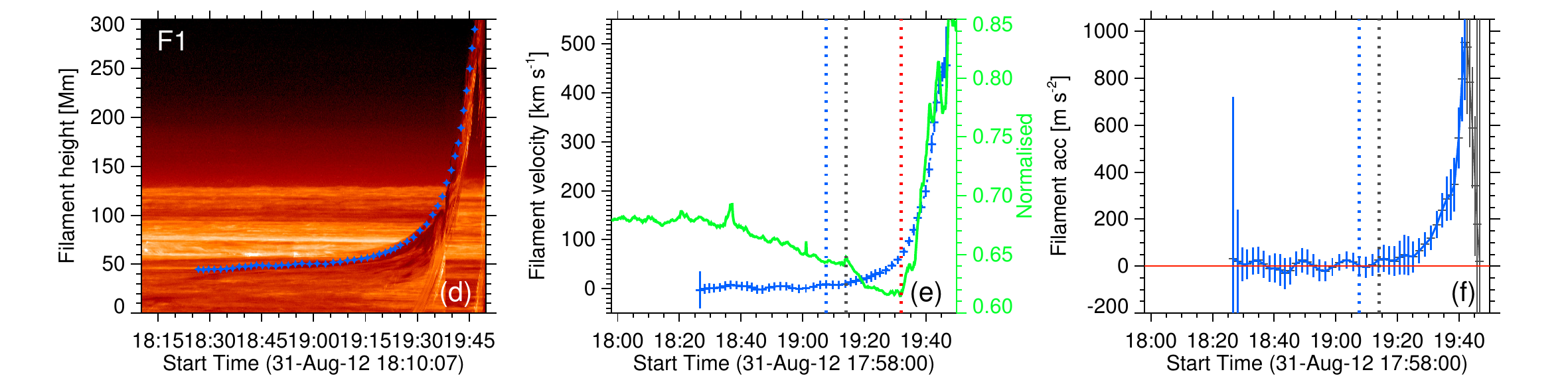}}
\caption{(a) Stack plot of the 131~{\AA} intensity along the dashed line in Figure~\ref{aia}(a). The diamonds show the measured height of the continuous moving front of H3. The arrow in yellow indicates the movement of nearby loops caused by the eruption of H3. (b) Temporal evolution of the velocity in the inner corona with vertical bars showing the uncertainty in velocity. The red curve shows the \textsl{GOES} 1--8~{\AA} SXR flux and the green curve shows the source-integrated AIA 131~{\AA} flux, which is normalized to its peak value. The flare onset ($t_{2}$) is indicated by the vertical line in red. The onsets of the main acceleration of H3 determined by the fitting method and from the acceleration-time profile are shown by the two vertical lines in blue ($t_{3}$) and \textbf{black} ($t_{4}$), respectively. (c) Temporal evolution of the acceleration with the data points in dark gray indicating a quickly decreasing acceleration. The horizontal line in red marks zero acceleration. (d)--(f) Same as (a)--(c) but for F1, with the green curve showing the normalized, source-integrated AIA 304~{\AA} flux.}
\label{ht}
\end{figure*}

\section{Results}\label{s:results}

\subsection{Early Kinematics of Solar Eruptions}\label{ss:kinematics}
\subsubsection{Temporal Evolution of Height, Velocity, and Acceleration}
We take the hot channel H3 and quiescent filament F1 as examples to illustrate our analysis procedure. Figures~\ref{aia}(a) and (b) display the 131~{\AA} image of H3 and the 304~{\AA} image of F1, respectively. From the supplemented movies, one can see that both H3 and F1 have a coherent structure and their fronts can clearly be identified throughout the eruption. Quiescent filaments and prominences are most likely trapped in dips of helical field lines, as indicated by the observations of cavities. Therefore, the magnetic axis of the erupting flux rope is best approximated by the upper edge of the erupting filament or prominence. For hot channels, the observations often suggest that they represent the erupting flux rope, whose magnetic axis should lie within the channel, roughly half way between the channel's upper and lower edges. However, the lower edge is often difficult to determine, as Figure~\ref{aia} illustrates. Therefore, we consider the upper edge of the hot channels to be the most reliable approximation of the erupting flux rope's magnetic axis.

In order to obtain the height of H3 and F1 vs. time, we make the time-slice plots as shown in Figures~\ref{ht}(a) and (d), respectively. The directions of the slices are chosen to ensure to cross the tops of H3 and F1 during most of the eruption, as indicated by the dashed lines in Figure~\ref{aia}. After inspecting all events in our sample, we find that their heights always present a two-phase evolution consisting of a slow-rise phase and a main-acceleration phase. Even for the failed H2 eruption, it still presents these two phases in spite of its short duration. We further examined the kinematics along different directions (within $5^\circ$ from the dashed lines Figure~\ref{aia}) and found that such a two-phase evolution patten always exists. In addition, we also inspect the influence of varying directions on the determination of the main-acceleration onset. This turns out to be smaller than that of varying the number of height-time data points, as illustrated in Section~\ref{ss:fit}. 

Taking advantage of the time-slice plots, we measure the projected heights of the H3 and F1 moving fronts as shown by the diamonds in Figures~\ref{ht}(a) and (d), respectively. Note that another moving feature appears slightly above the upper edge of H3 during the slow-rise phase (as shown by the yellow arrow in Figure~\ref{ht}(a)). This is a separate structure consisting of nearby loops that were produced by a previous confined flare in the same region (see the attached movie). In order to derive the velocity, we apply the first order numerical derivative routine \texttt{deriv.pro} from the IDL software package to the smoothed height-time (H-t) data, with a cubic spline smoothing performed by the IDL routine \texttt{IMSL\_cssmooth.pro} to reduce the noise. The results are shown in Figures~\ref{ht}(b) and (e). Using the second order numerical derivative, we further derive the acceleration as shown in Figures~\ref{ht}(c) and (f). The uncertainties in velocity and acceleration mainly stem from the uncertainty of the measured heights, which is estimated to be 2 (3)~pixels for the hot channels (filaments).

\begin{figure*}
 \center {\includegraphics[width=12cm]{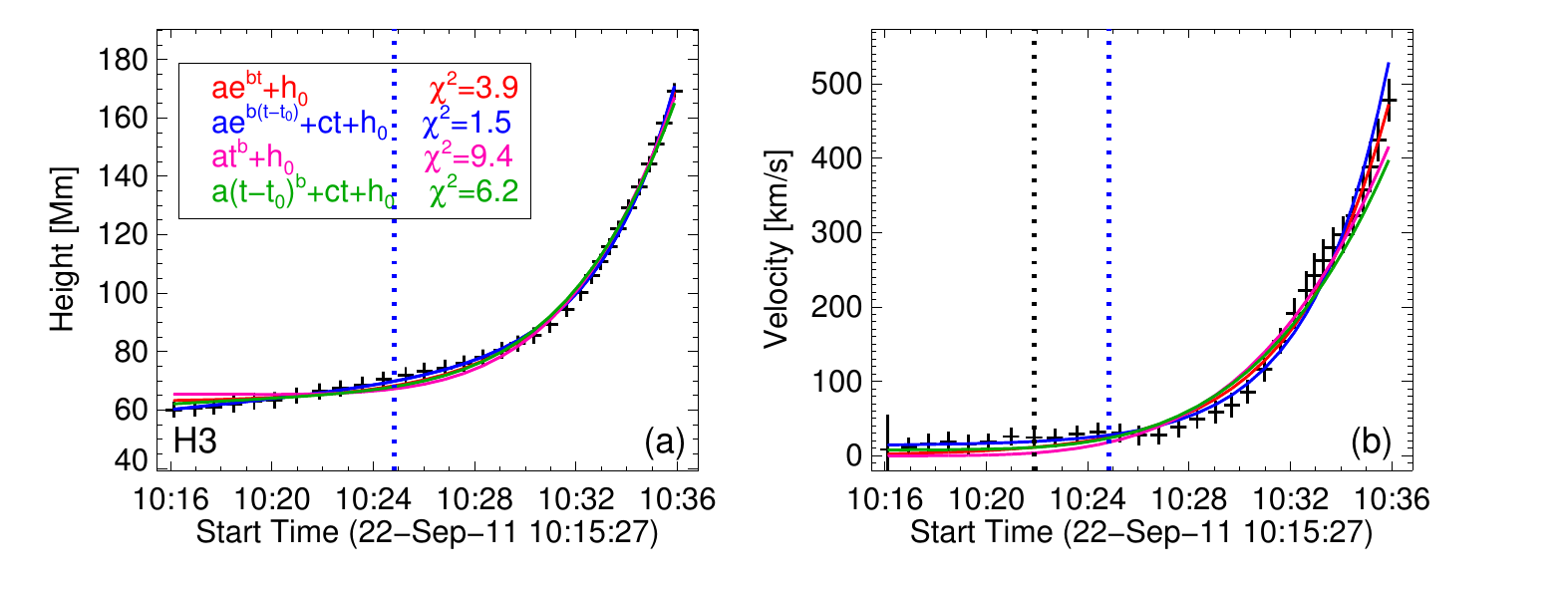}}
 \center {\includegraphics[width=12cm]{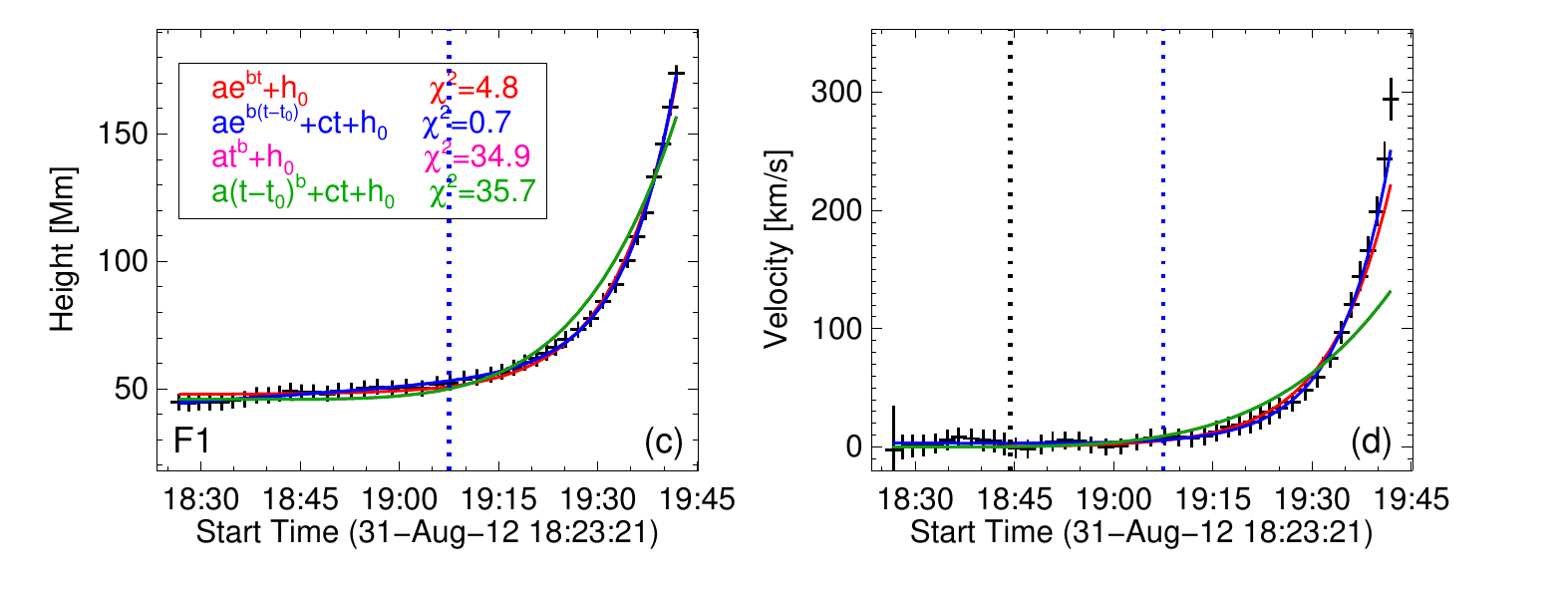}}
\caption{Model fitting of the rise profile for H3 (a--b) and F1 (c--d). The lines in green, red, blue, and pink show the fitting results of the four functions as shown in the top left corner of Panels (a) and (c). The vertical lines in blue indicate the onset time $t_3$ of the main-acceleration obtained from the best fit of $h_2(t)$ with prescribed $t_0=0$, and the vertical lines in black show the resulting $t_0$ when it is included as a free parameter, $t_0\ge0$, in the fit.}
\label{fit}
\end{figure*}

\begin{table*}
\caption{Metrics for fitting goodness of different functions.}
\label{tb2}{
\hspace{0.1\textwidth}
\begin{tabular}{ccccc}
\\ \tableline \tableline
 Events & a$_{1}$e$^{b_{1}t}$+h$_{0}$   & a$_{2}$e$^{b_{2}(t-t_0)}$+c$_{2}$t+h$_{0}$ & a$_{3}$t$^{b_{3}}$+h$_{0}$ & a$_{4}(t-t_{0})^{b4}$+c$_{4}$t+h$_{0}$    \\
             &$\chi_{\nu1}^{2}$  &$\chi_{\nu2}^{2}$ & $\chi_{\nu3}^{2}$ &$\chi_{\nu4}^{2}$\\
\tableline 
H1     & 3.7     & 0.5     & 12.7     & 1.0  \\
H2     & 10.6   & 0.9     & 16.1     & 5.4  \\
H3     & 3.9    &1.5       & 9.4       & 6.2  \\
H4     & 37.0  &1.7       & 62.6    & 17.3  \\
H5     & 2.6    & 2.8      & 9.5      & 1.8  \\
H6     & 3.5    & 3.8      & 1.7      & 1.5  \\
\tableline
F1     &4.8       &0.7      &34.9        &35.7  \\
F2     &4.5       &1.4      &7.6          &9.1  \\
F3     &3.7       &0.5      &6.0          &3.7  \\
F4     &14.5     &1.6      &29.2        &33.1  \\
F5     &42.2     &1.5      &926.3      &949.8  \\
F6     &7.8       &0.9      &18.8        &21.7  \\
\tableline
\end{tabular}}\\
\end{table*}

Figure~\ref{ht}(b) shows that, during the slow-rise phase, the H3 eruption has a small, weakly accelerated rise velocity in the range 10--30~km~s$^{-1}$. After serval minutes, it starts to speed up. The velocity increases from about 50~km~s$^{-1}$ to 600~km~s$^{-1}$ in only 8~min, corresponding to an average acceleration of 1150~m~s$^{-2}$. The temporal variation of the acceleration (Figure~\ref{ht}(c)) shows that the acceleration is very small during the slow-rise phase. It starts to increase strongly at $\sim$10:26~UT, peaks at $\sim$10:36~UT, and then decreases quickly. Such a quick decrease may be due to a decreasing visibility of the structure above the height of $\sim$170 Mm ($\sim$200 arcsecs above the solar limb), which is typical for hot channel eruptions, resulting in a substantial underestimate of the height. The other potential reason is a beginning saturation of the instability that drives the eruption, manifesting as a decrease of the acceleration. 

Figures~\ref{ht}(d)--(f) show that the F1 eruption has similar height-time, velocity-time, and acceleration-time profiles to that of H3, however, the duration is much longer. It takes $\sim$40~min for the weakly accelerated slow rise to reach a similar velocity of $\sim$30~km~s$^{-1}$. During the main-acceleration phase, the velocity varies relatively more slowly, increasing from $\sim$30~km~s$^{-1}$ to $\sim$450~km~s$^{-1}$ in $\sim$30~min with an average acceleration of about 230~m~s$^{-2}$. Figure~\ref{ht}(f) shows that the acceleration of F1 is also centred around zero, and then starts to increase rapidly, followed by a decrease when approaching the limit of the AIA field of view. It is worth mentioning that the projection effect has a significant influence on the measured heights, velocities, and accelerations of the quiescent filaments but not on the character of their temporal profiles. We will estimate the true heights in Section~\ref{ss:TI}. Moreover, for both hot channels and quiescent filaments, the solar rotation has some contribution to the velocity, which is, however, very small ($<$1~km~s$^{-1}$) and can be neglected \citep{mcCauley15}.

\subsubsection{Fit of Height-time Profiles}\label{ss:fit}

In order to infer the functional forms of the slow-rise and main-acceleration phases and a possible break point between them, we consider a set of fit functions for the measured height-time profiles. A nonlinear function is required to fit the main-acceleration phase; here we include the exponential and the power law, as suggested by previous work (see Section~\ref{s:intro}). We do not include the often used tanh function \citep[e.g.,][]{sheeley07} because this extends the fitting into the propagation phase after the main acceleration, which is beyond the scope of the present investigation. Moreover, a linear or quadratic function appears appropriate for the slow-rise phase, as the acceleration in this phase is typically \emph{much} weaker (Figure~\ref{ht}). A constant term includes the initial height for each event. In order to determine whether a break point exists, we compare the nonlinear fit functions with a superposition of the nonlinear and the linear or quadratic functions. The superposition should yield the better fit in the presence of a break point. The application of each fit function to the whole time series for each event ensures comparability between the fits, because the measure of goodness, the reduced chi-squared, $\chi_\nu^2$, is then based on the same number of data points for each fit function.

The fit is performed for the main part of the height-time profiles including all measured heights up to the final point of increasing acceleration. This is consistent with the character of all fit functions, which do not include a decreasing second derivative. In trying to fit the superposed functions to the data, it is found that the fitting software often cannot find a better fit when the quadratic term is included. In some cases, a poorer fit than that excluding the quadratic term is obtained, although a vanishing coefficient for this term would be a valid solution, providing a very similar goodness of fit as the superposition with the linear function. In the interest of using a uniform method for all events, we have, therefore, dropped the quadratic term. This aspect is elaborated further by applying the quadratic fit function (including the linear term) only to the slow-rise phase that is inferred from the superposed fit. It turns out that the uncertainty of the acceleration is bigger than, or comparable to, the inferred acceleration for 6 of the 12 events (see detail below). This additionally suggests to restrict the fitting of the slow-rise phase to a linear function, although the velocity data show a small increase in this phase for several of our events. Thus, the following functions are employed in the fitting, 
\begin{subequations}
\begin{align}
   h_1(t) &= a_1 \exp(b_1t)                + d_1\,, \\ 
   h_2(t) &= a_2 \exp[b_2 (t-t_0)] + c_2 t + d_2\,, \\ 
   h_3(t) &= a_3 t^{b_3}                   + d_3\,, \\ 
   h_4(t) &= a_4 (t-t_0)^{b_4}     + c_4 t + d_4\,, 
\end{align}
\end{subequations}
where $h_i$ and $t$ denote fitting height and time, respectively. The quantities $a_i$, $b_i$, $c_i$, $d_i$, and $t_0$ are the coefficients of the functions to be determined by the fit. The fit is performed by the routine \texttt{mpfit.pro} \citep{Markwardt09}, which is available in the Solar SoftWare (SSW) package. The reduced chi-square $\chi_\nu^{2}$ is calculated by $\frac{\chi^{2}}{N-m}$, where $\chi^{2}=\sum_{i=1}^{i=N} \frac{[h_i(t)-H_i(t)]^{2}}{\sigma_i^{2}}$, $N$ denotes the number of data points, $N-m$ is the number of the degrees of freedom, i.e., number of data points minus the number of free parameters ($m$) in the fit function, and $\sigma_i$ is the error for each measured height $H_i(t)$. The best fit is indicated when $\chi_\nu^{2}$ is closest to unity. 

For the nonlinear component of the superposed functions, we employ a two-step strategy. First, to minimize the difference between the purely nonlinear and the superposed fit functions, we set $t_0=0$. This allows us to compare the resulting fits on the formally most equal basis, but implies the assumption that the nonlinear evolution of the rise commences simultaneously with the slow rise (our first data point). Since we also intend to address the question whether the nonlinear evolution (indicating onset of instability) starts associated with the slow rise or with the main acceleration, we treat $t_0$ as a free parameter in a second step.

\begin{figure*}
 \center {\includegraphics[width=9cm]{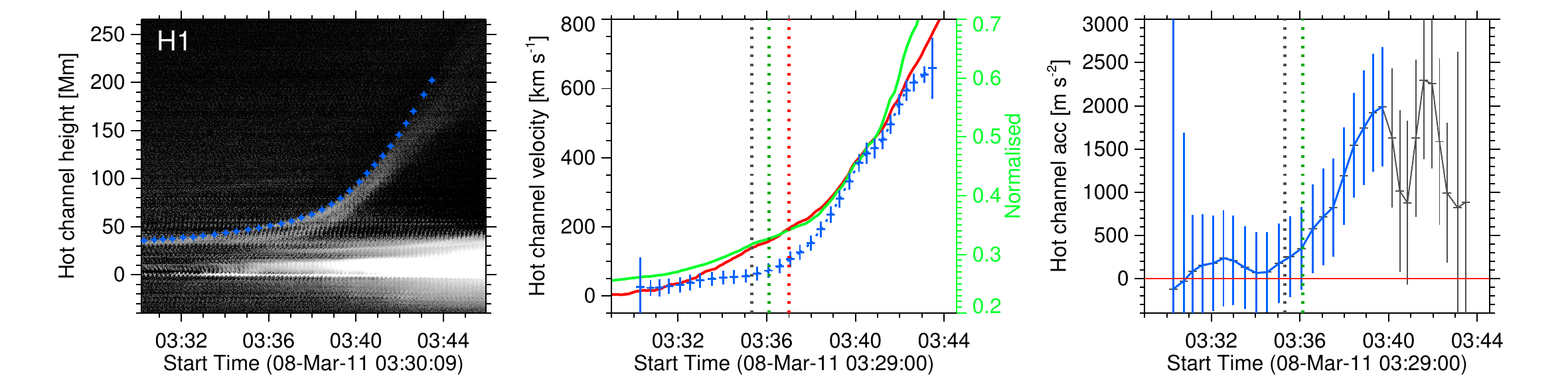}\includegraphics[width=6cm]{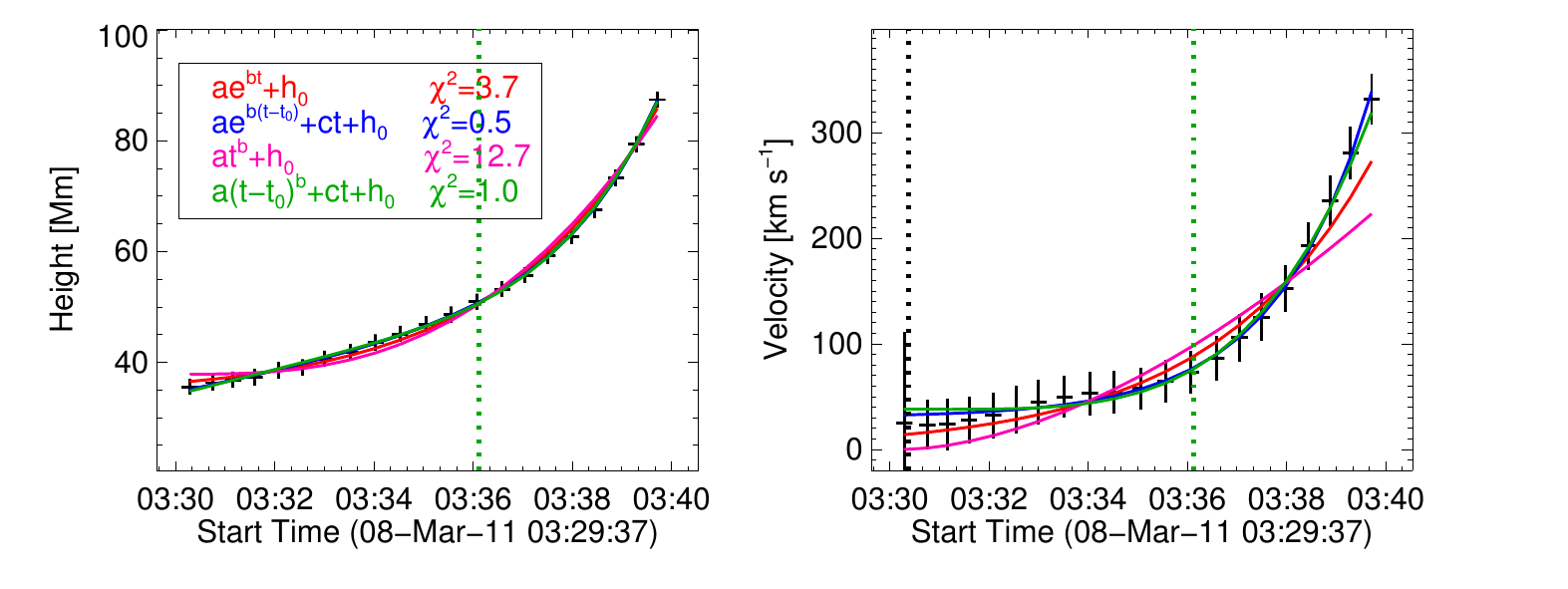}}
 \center {\includegraphics[width=9cm]{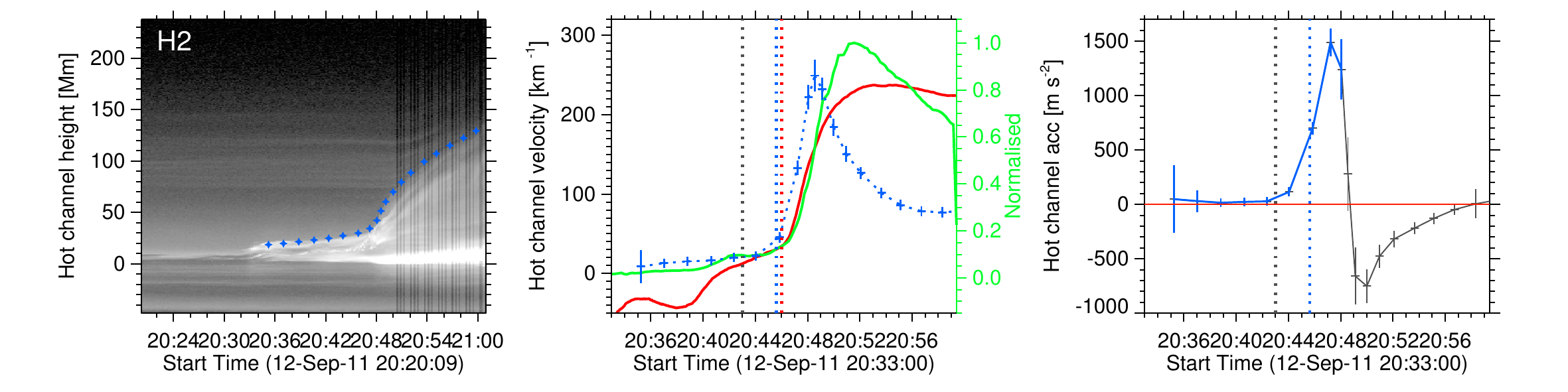}\includegraphics[width=6cm]{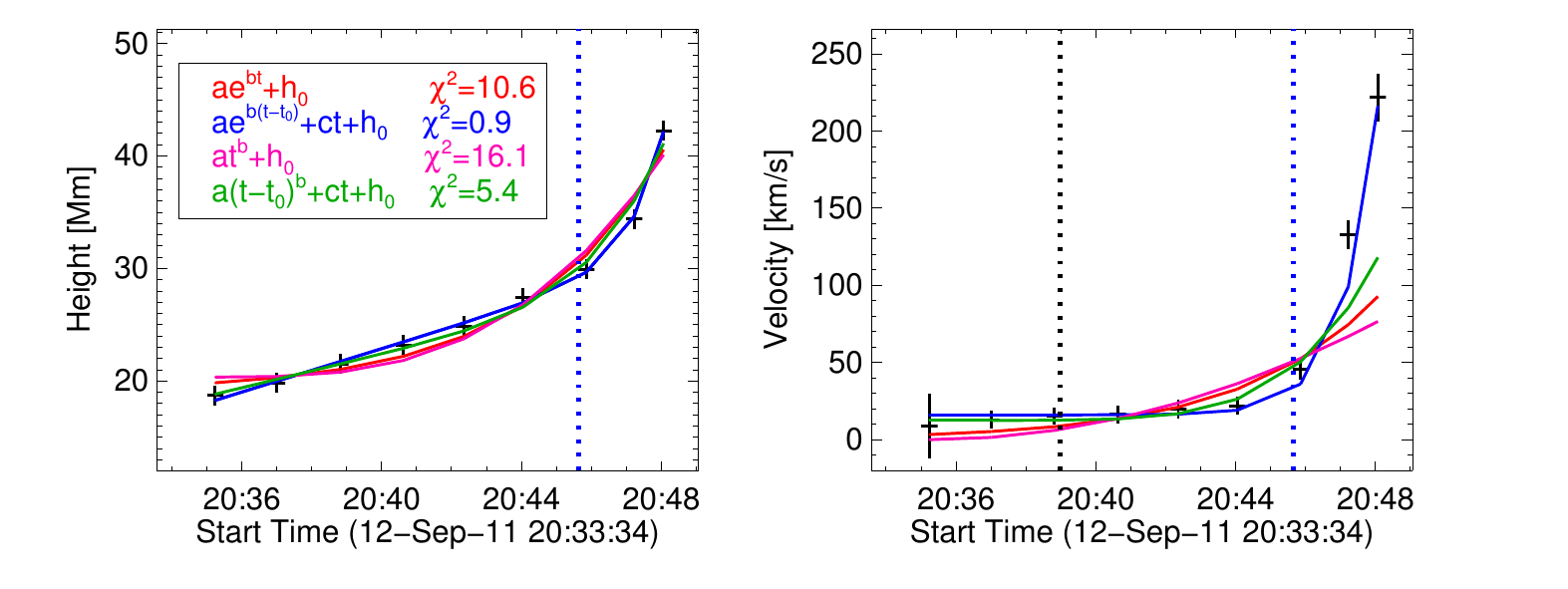}}
 \center {\includegraphics[width=9cm]{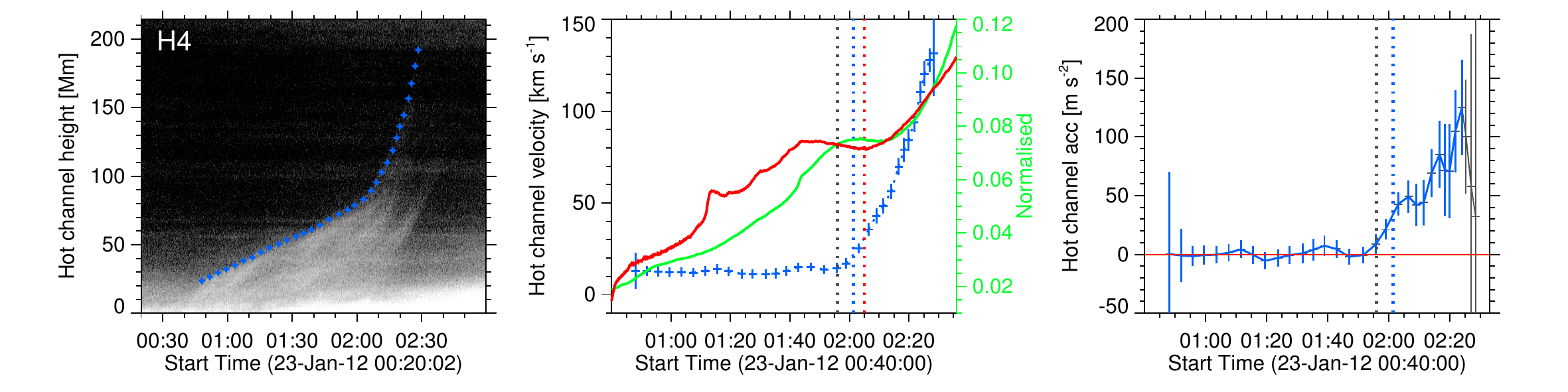}\includegraphics[width=6cm]{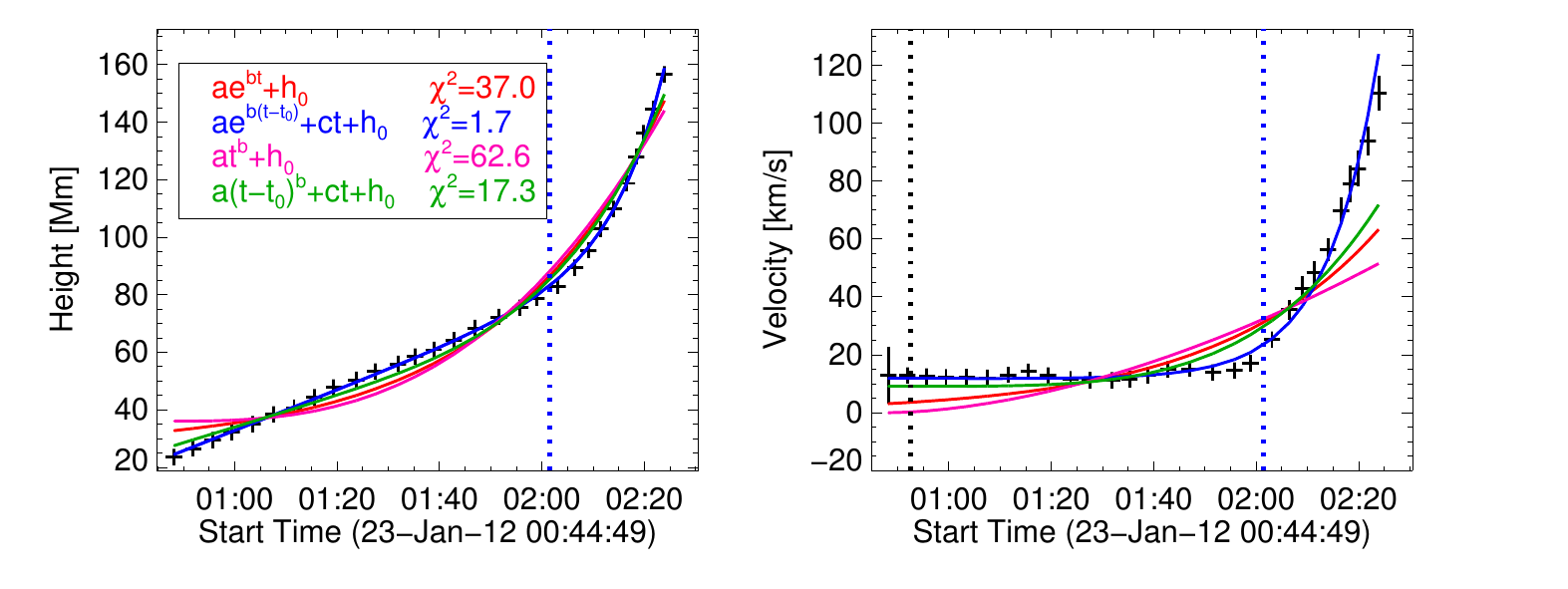}}
 \center {\includegraphics[width=9cm]{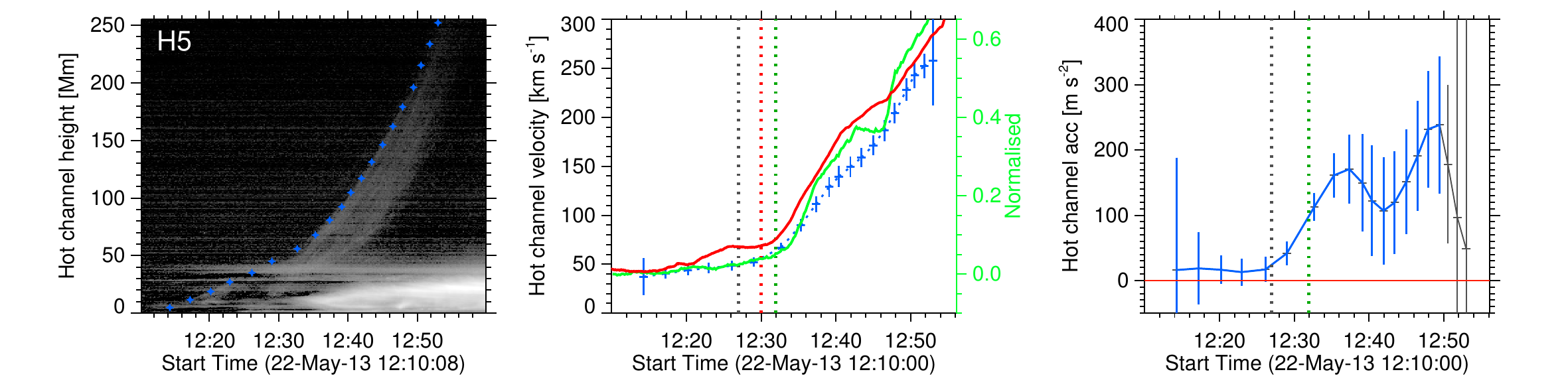}\includegraphics[width=6cm]{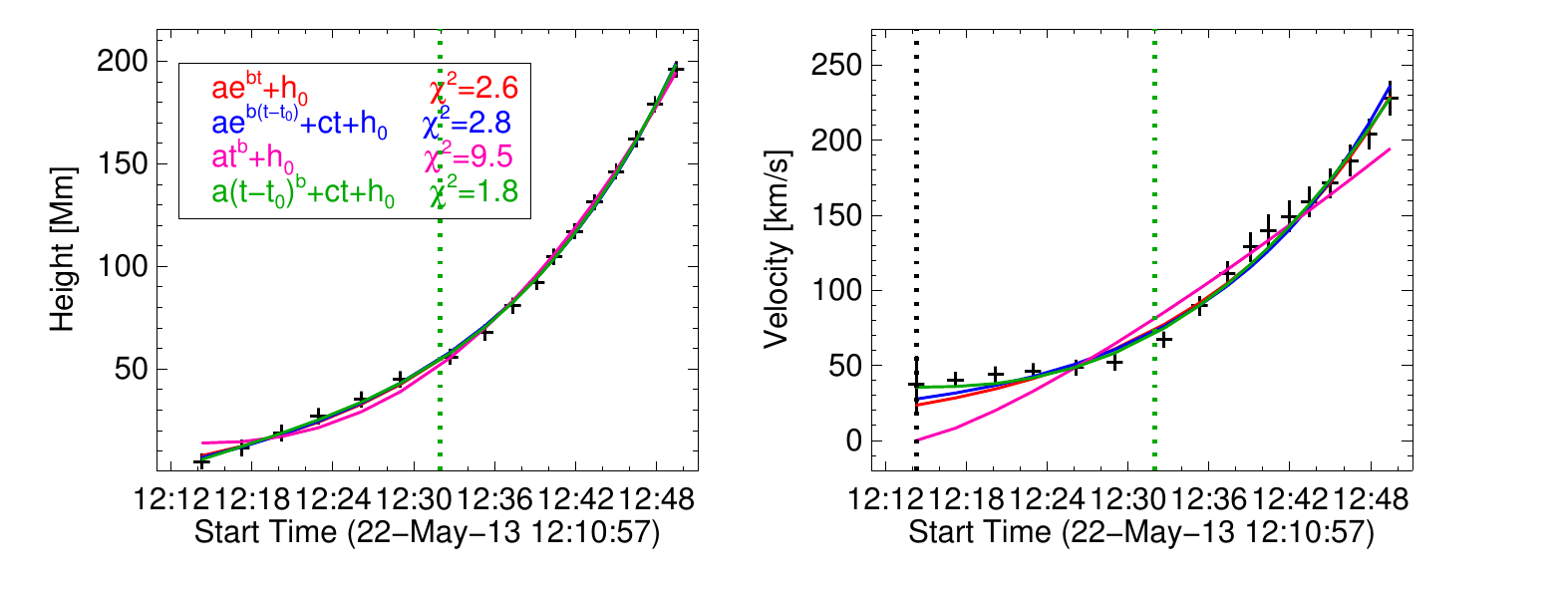}}
 \center {\includegraphics[width=9cm]{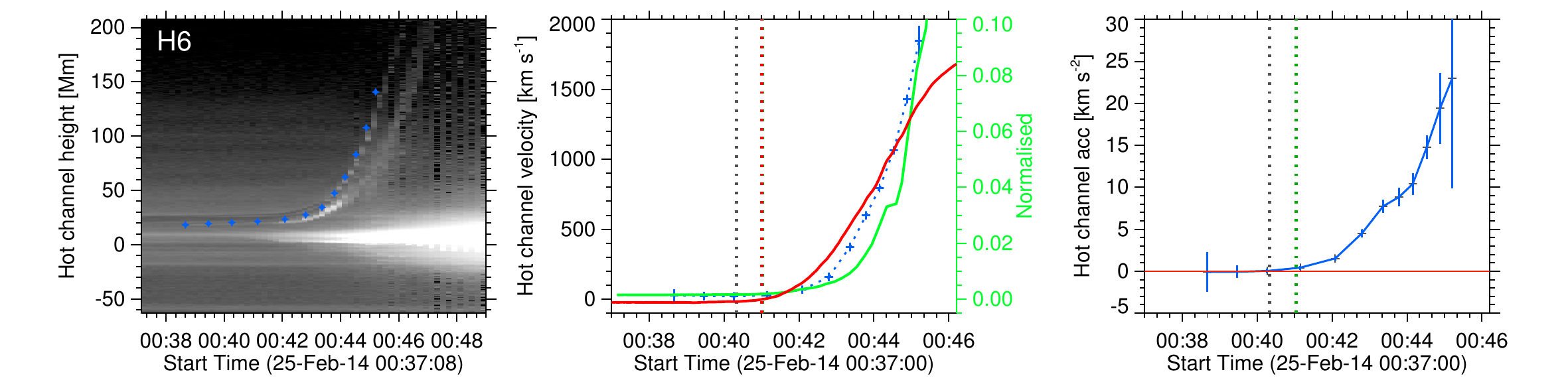}\includegraphics[width=6cm]{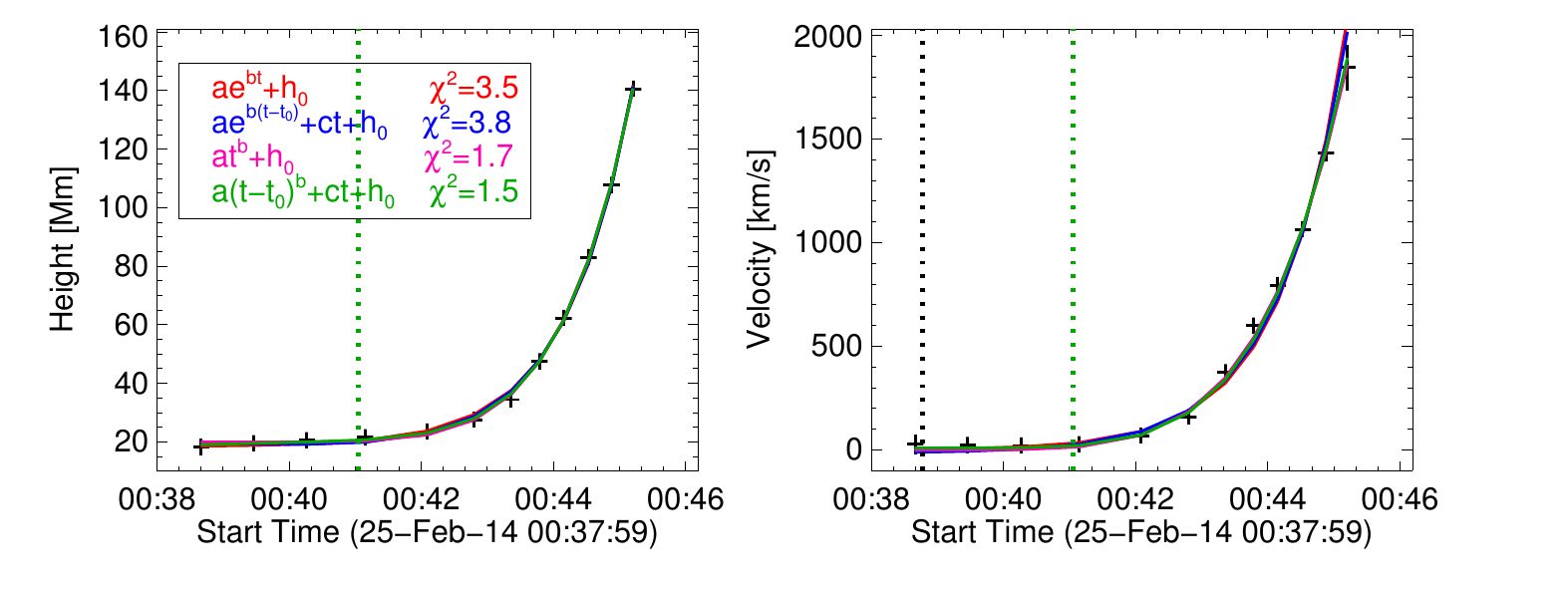}}
\caption{Same as Figures~\ref{ht} and \ref{fit} but for H1, H2, and H4--H6.}
\label{ht1}
\end{figure*}

\begin{figure*}
 \center {\includegraphics[width=9cm]{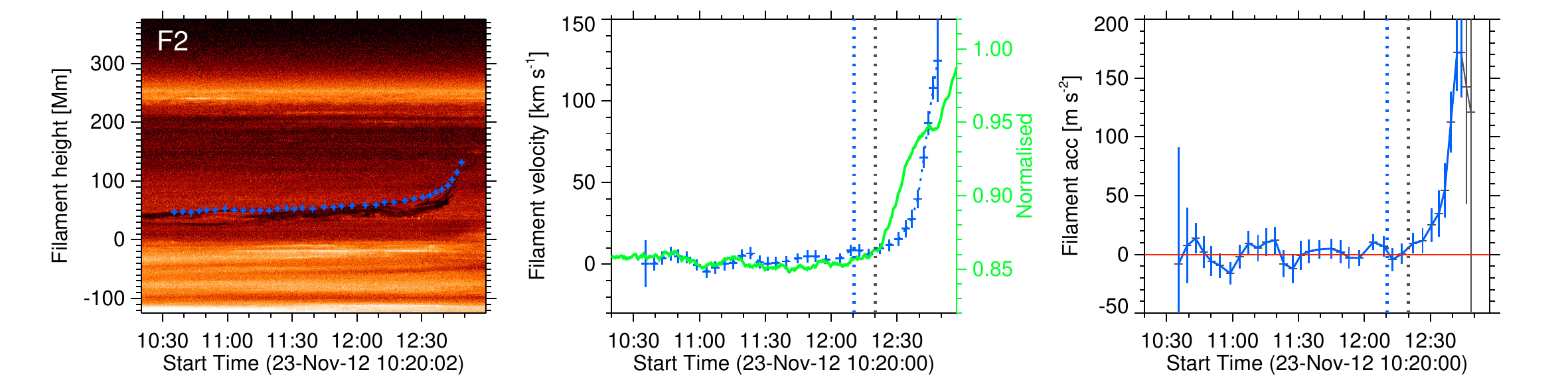}\includegraphics[width=6cm]{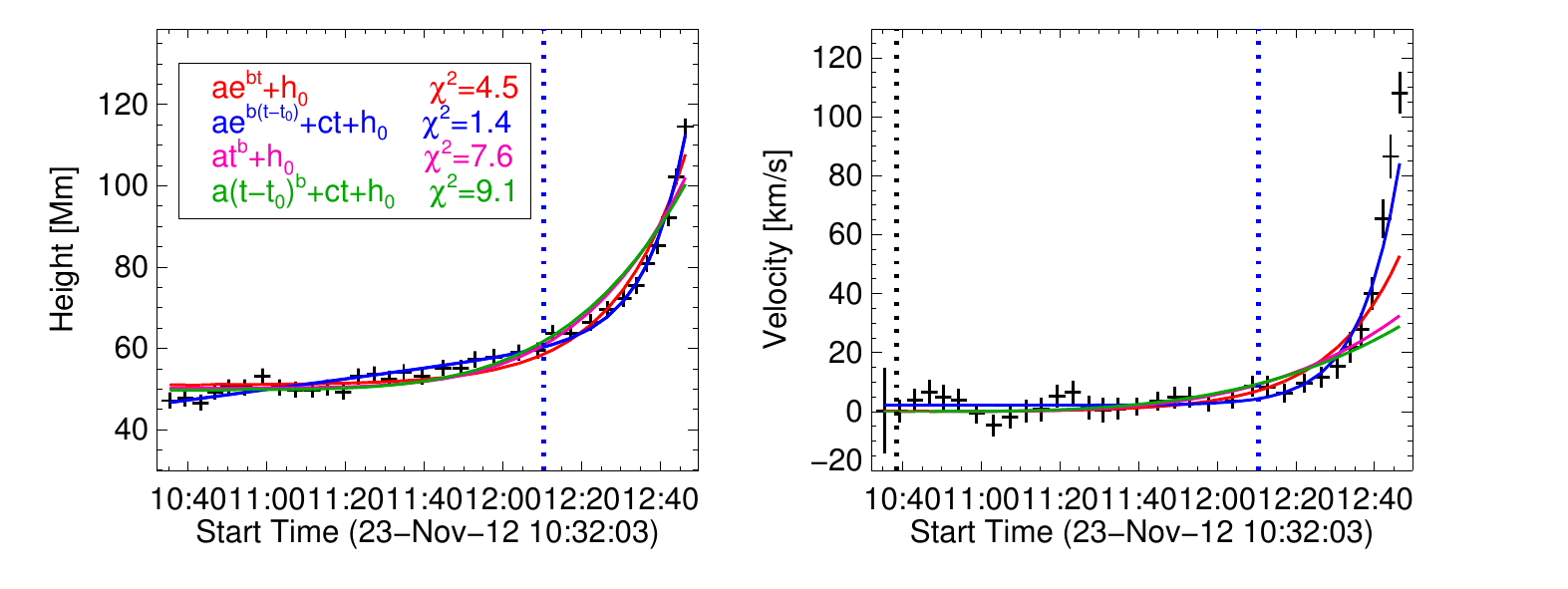}}
 \center {\includegraphics[width=9cm]{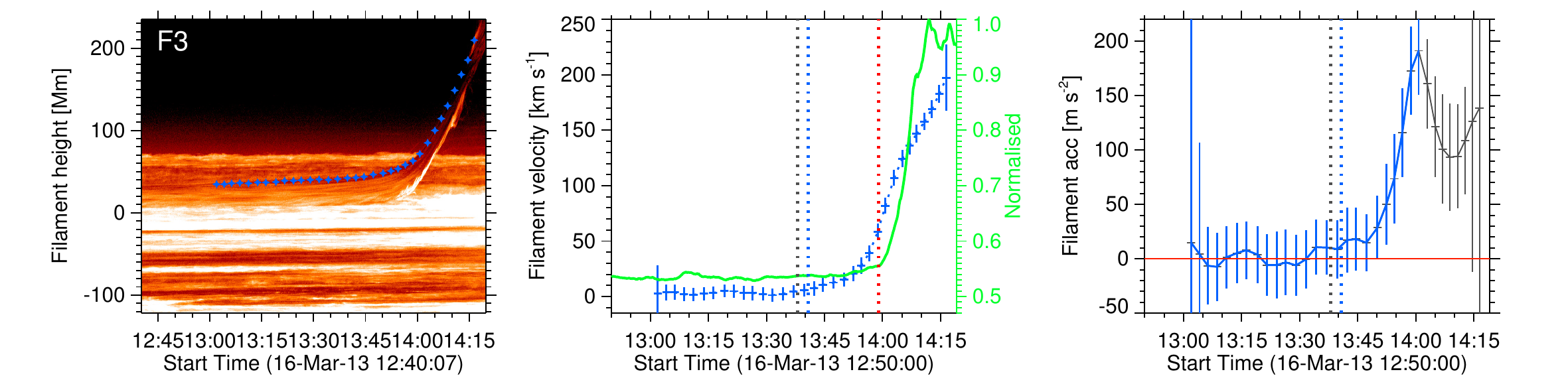}\includegraphics[width=6cm]{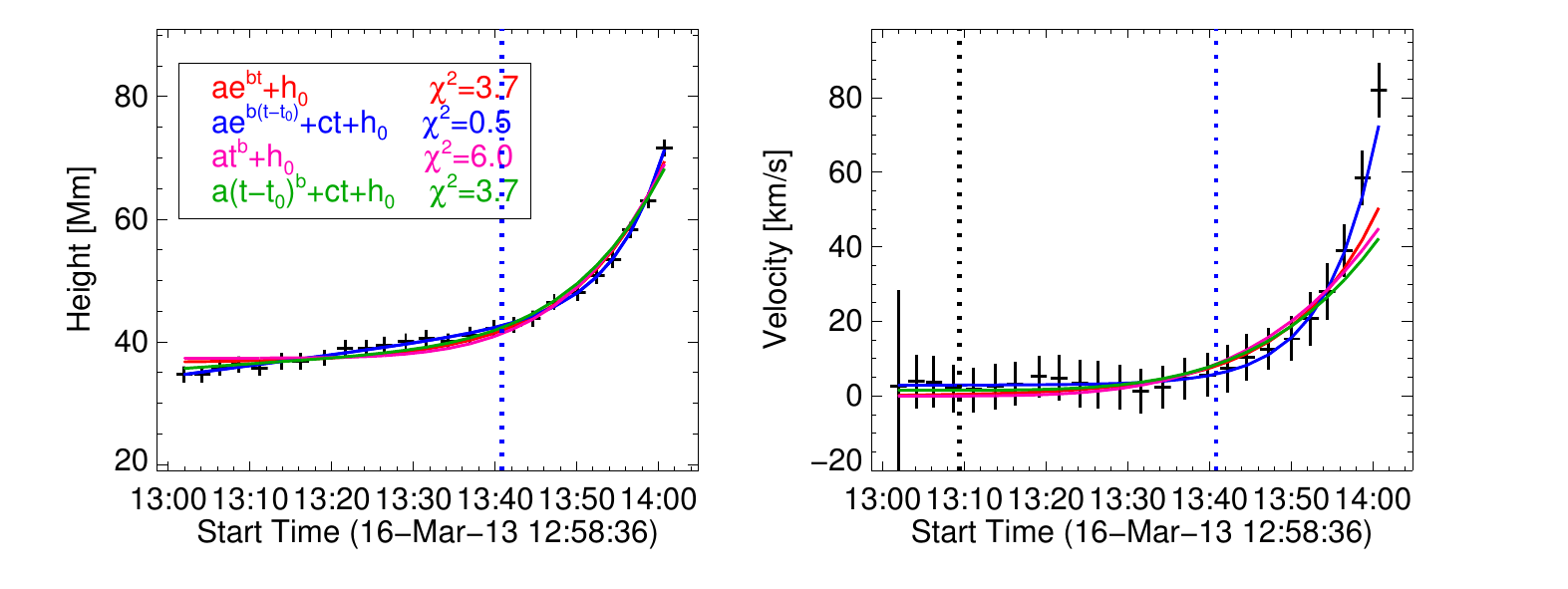}}
 \center {\includegraphics[width=9cm]{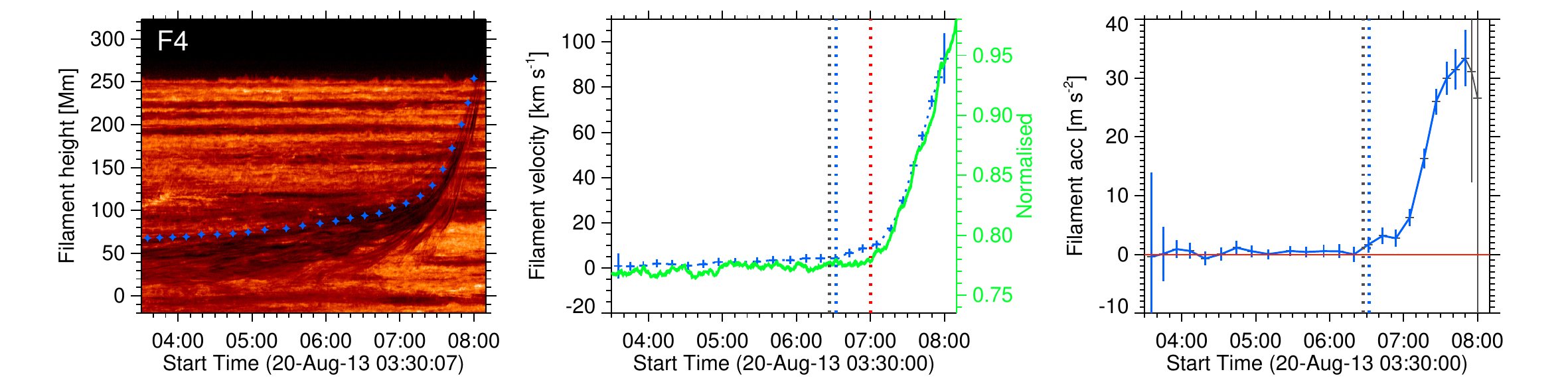}\includegraphics[width=6cm]{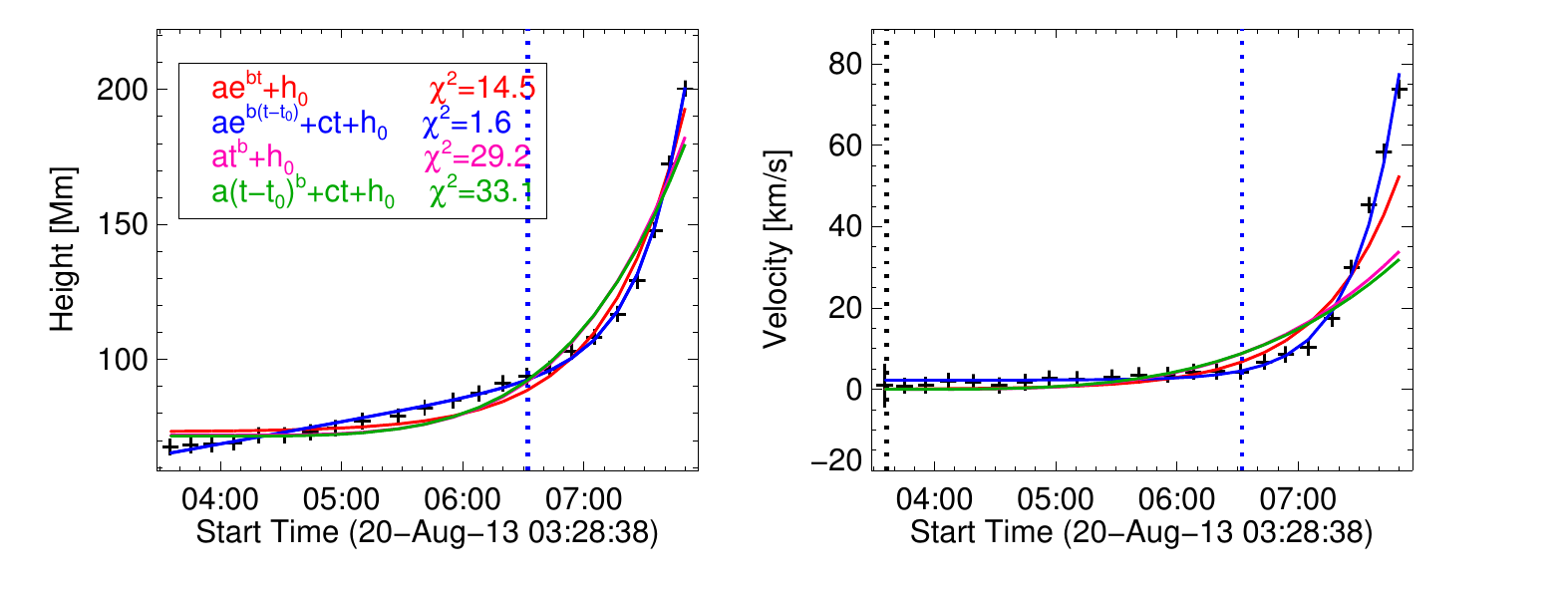}}
 \center {\includegraphics[width=9cm]{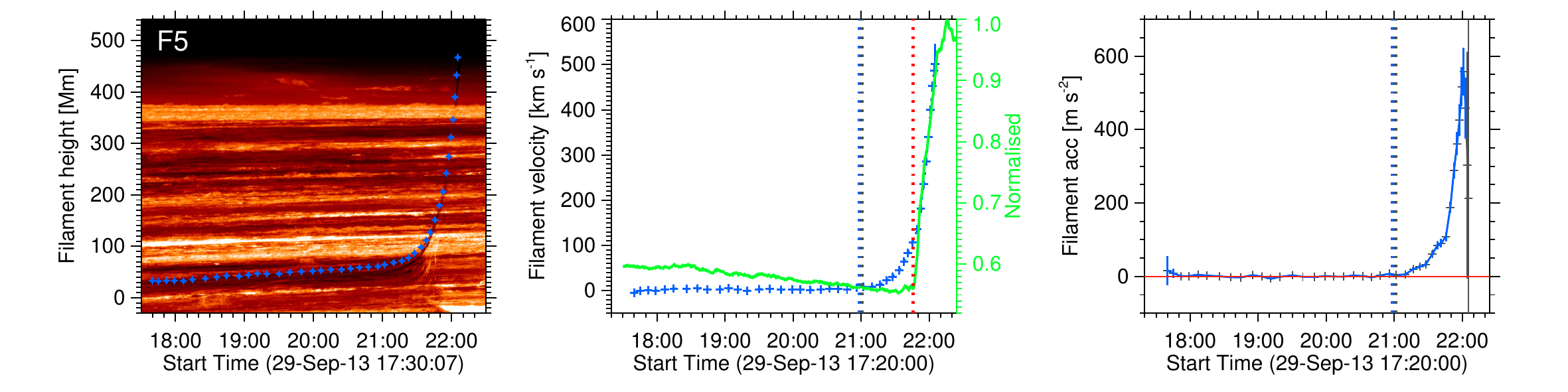}\includegraphics[width=6cm]{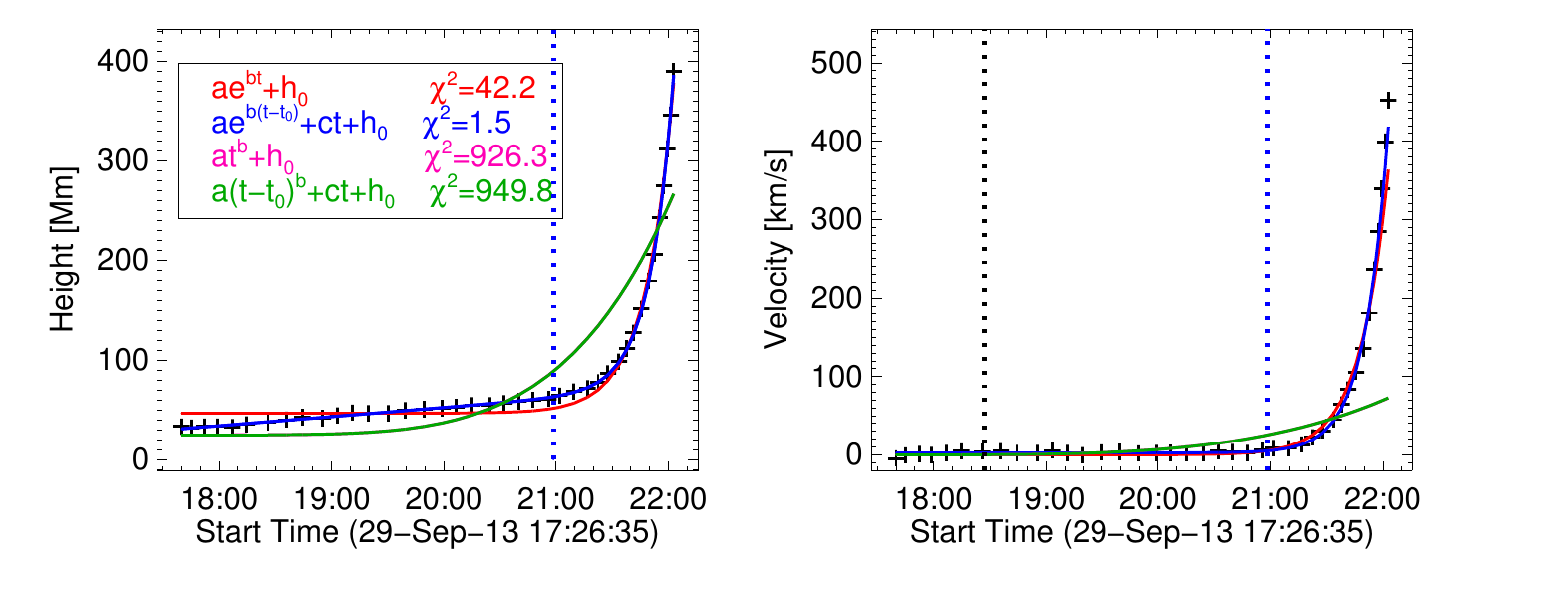}}
 \center {\includegraphics[width=9cm]{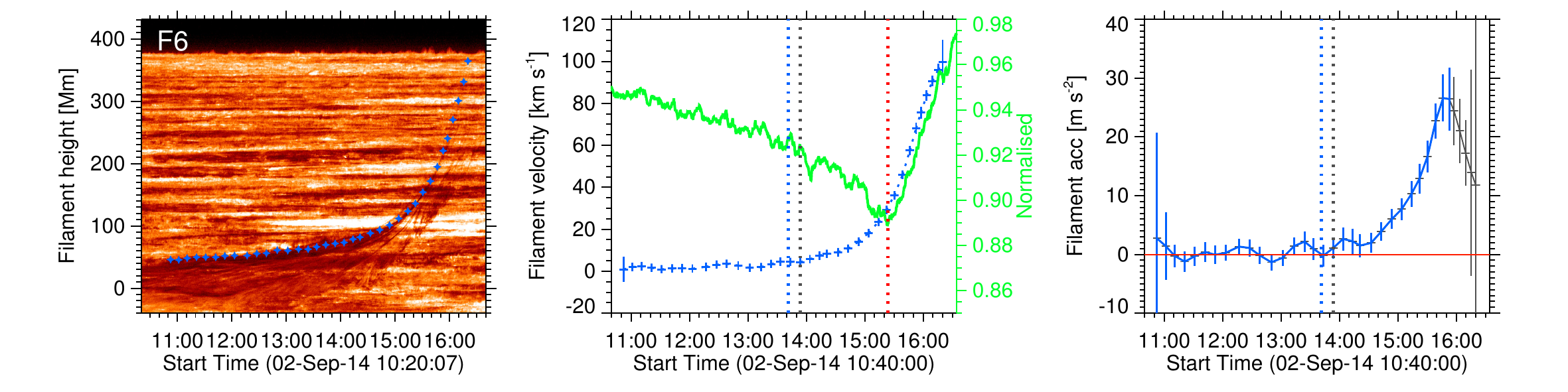}\includegraphics[width=6cm]{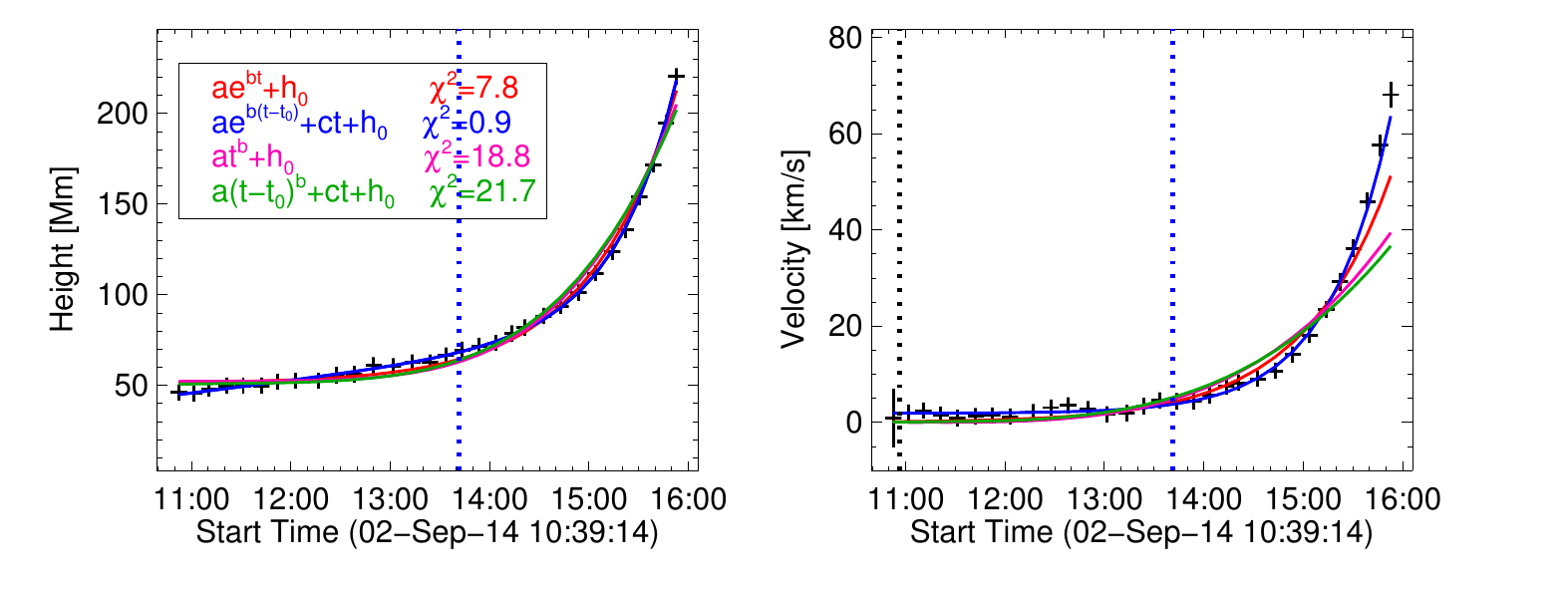}}
\caption{Same as Figures~\ref{ht} and \ref{fit} but for F2--F6.}
\label{ht2}
\end{figure*}

The results for H3 and F1 are shown in Figure~\ref{fit}. One can see that all functions can fit the height-time profiles relatively satisfactorily, but the goodness of fit can be obviously distinguished in the velocity-time profiles. This is also apparent for most of the other events, shown in Figures~\ref{ht1} (hot channels) and \ref{ht2} (quiescent filaments). The existence of a different functional form, hence the existence of a break point, between the slow-rise and main-acceleration phases is demonstrated by the clear superiority of the $\chi_\nu^{2}$ values for one or both of the superposed functions $h_2(t)$ and $h_4(t)$ for all 12 events (Table~\ref{tb2}). Additionally, the exponential fit is superior to the power-law fit for the majority of our events (three of six hot channel eruptions and all six quiescent filament eruptions). For each event, we obtain an estimate of the break-point time where the velocity of the nonlinear term starts to take over (equals) that of the linear term in the best-fitting superposed function, $h_2(t)$ or $h_4(t)$. This time is given as $t_{3}$ in Table~\ref{tb3}. It has an uncertainty which we estimate by excluding a varying number of the first or final data points from the best fit. The uncertainty of the break point is estimated to be within $\sim$4~min for the hot channels and $\sim$10~min for the quiescent filaments. 

Next, the superposed fits are repeated allowing for a non-zero onset time of the nonlinear evolution, $t_0\ge0$. It turns out that the shape and $\chi_\nu^{2}$ value of the best fit are nearly identical to the fit with prescribed $t_0=0$, but most resulting fit values for the onset time $t_0$ strongly precede the break-point times $t_{3}$, especially for the quiescent filaments; see Table~\ref{tb3}. Comparing the inferred value for $t_0$ with the shape of the velocity-time profile, it is seen in Figures~\ref{fit}--\ref{ht2} that $t_{3}$ appears to be the far more reliable estimate of the onset of the main acceleration than $t_0$ for most events. Therefore, we will not use the inferred onset times $t_{0}$ in the further analysis. One should be aware that the main acceleration may start somewhat earlier than $t_{3}$ because it needs some time to build up a velocity comparable to the slow-rise velocity. For a rough estimate of this time, we adopt the assumption that the main acceleration starts from near the time $t_{s}$ where the nonlinear velocity component satisfies $v(t_{s})=v(t_{3})/10$; the value of $t_{s}$ is also listed in Table~\ref{tb3}.  

\begin{figure}
\center {\includegraphics[width=8cm]{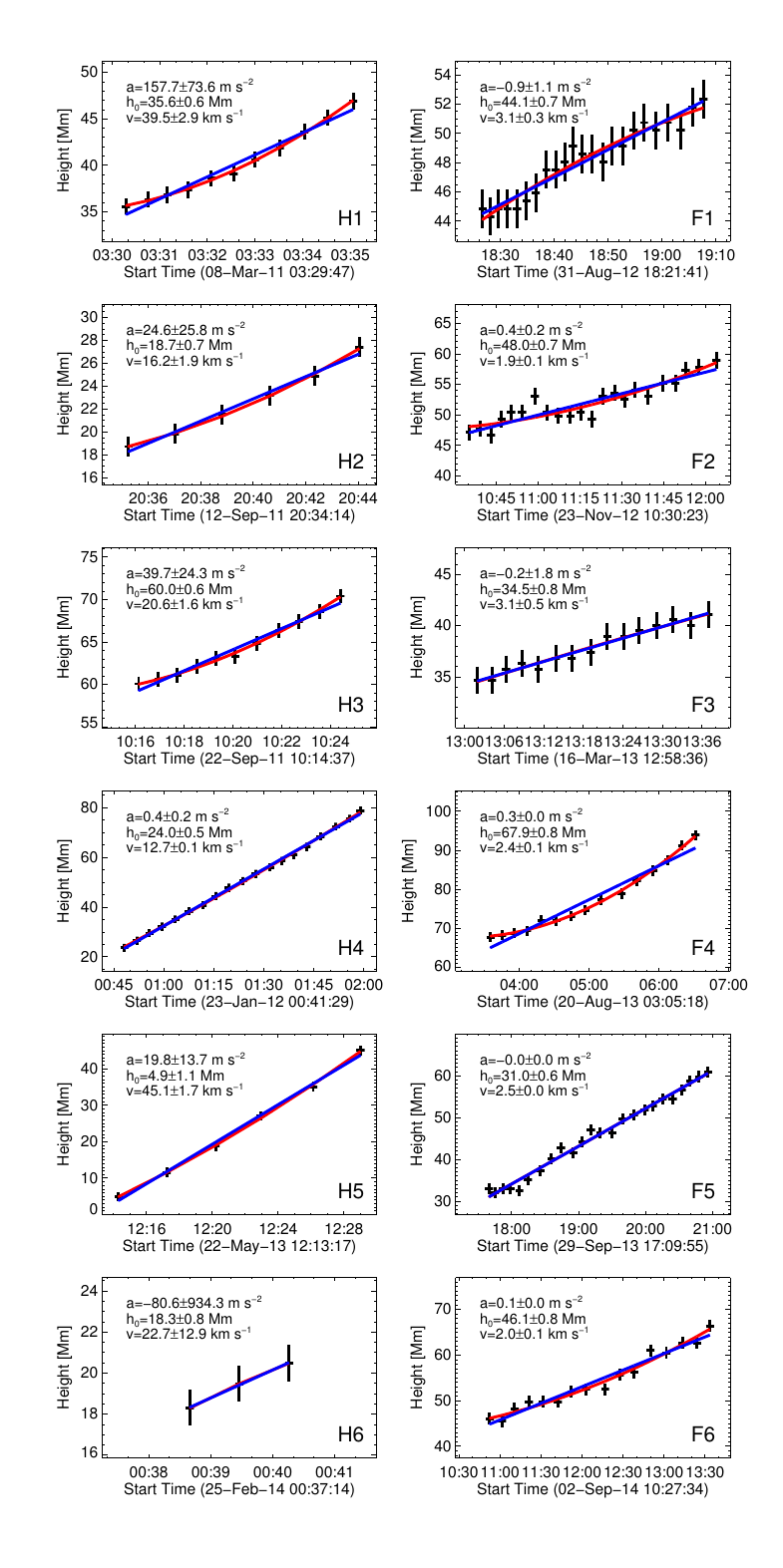}}
\caption{Linear (blue) and linear plus quadratic (red) fit to the slow-rise phase of all eruptions.}
\label{slowfit}
\end{figure}

In order to further validate our judgment with regard to excluding the quadratic term, we only fit the data points in the slow-rise phase, before the inferred break point $t_3$, with the linear and linear-plus-quadratic functions. The results are shown in Figure~\ref{slowfit}. One can see that the linear fit is generally very close to the quadratic one. Only H1, H3, and H5 show a significant average acceleration, which is moderate only for H1, otherwise small. H4, F2, F4, and F6 also obtain a valid acceleration, however, the value is extremely small ($<$0.5~m~s$^{-2}$). For the other events, the uncertainty of the inferred acceleration is bigger than the acceleration itself, which appears to be due to a small number of data points (H2, H6) or an oscillatory behavior of the slow-rise velocity and acceleration (F1, F3, F5). 

From the break point $t_3$ and our first data point for the slow-rise phase, we obtain its duration, which is shown by $D$ in Table~\ref{tb3} and Figure~\ref{para1}(a). One can see that the duration of the slow-rise phase for CMEs from the hot channel eruptions is mostly much shorter than that of the erupting filaments. The latter all have a slow rise of $>$40~min and up to $>$170~min for F4--F6. The H4 event also has a slow-rise phase of long duration ($\sim$75 min). As noted already above, a long and high hot channel connecting the periphery of two active regions erupts in this event, so that the corresponding magnetic field strength is much weaker than that of the other five hot channels which originate from the central area of their active regions. This resembles the quiescent filaments which originate from large-scale and weak magnetic flux.

\subsection{Timing Relation to Flares}\label{ss:timing}
\subsubsection{Main-acceleration Phase}
We compare the early kinematic evolution of the eruptions with the evolution of the associated flare emission. Figure~\ref{ht}(b) shows that the velocity evolution of H3 is closely synchronized with the \textsl{GOES} 1--8~{\AA} SXR light curve during the main part of the main-acceleration phase. Since the \textsl{GOES} flux is from the full disk and may include a contribution from other regions, we also integrate the AIA 131~{\AA} intensity in the H3 source region to represent the flare emission of the event. The velocity evolution is even somewhat better synchronized with the evolution of the integrated AIA 131~{\AA} flux. Except H4, such a simultaneity is also true for the other hot channel events although not always tightly close (e.g., the H1 event, Figure~\ref{ht1}). The exception of H4 is mainly due to the fact that it originates from a large-scale magnetic structure which connects two nearby active regions, whereas the \textsl{GOES} flux is from two successive flares occurring at two different locations in the active-region complex \citep{cheng13_double}. 

For the quiescent filaments, we utilize the 304~{\AA} intensity as the proxy of the associated flare emission because the relevant ribbons and arcades only clearly appear in the AIA low-temperature passbands but not in SXRs and AIA high-temperature passbands. These flares are usually very weak, with the plasma not being heated above a temperature of $\sim$10~MK (the peak temperature of the 131~{\AA} response function). We also inspect the 171~{\AA} and 211~{\AA} fluxes and find that their profiles are very similar to that of the 304~{\AA} flux. Similar to the hot channels, the velocity evolution of the erupting filaments keeps in step with the evolution of the integrated AIA 304~{\AA} flux in the main-acceleration phase (Figures~\ref{ht}(e) and \ref{ht2}). Again, the synchronization is not tight in some events (F2 and F3). Overall, this indicates that the mechanism of the main CME acceleration and the mechanism of the rapid increase of the flare emission are coupled, often closely, for both hot channels and quiescent filaments during the main part of the energy release in the eruptions.

\begin{figure}
 \center {\includegraphics[width=8cm]{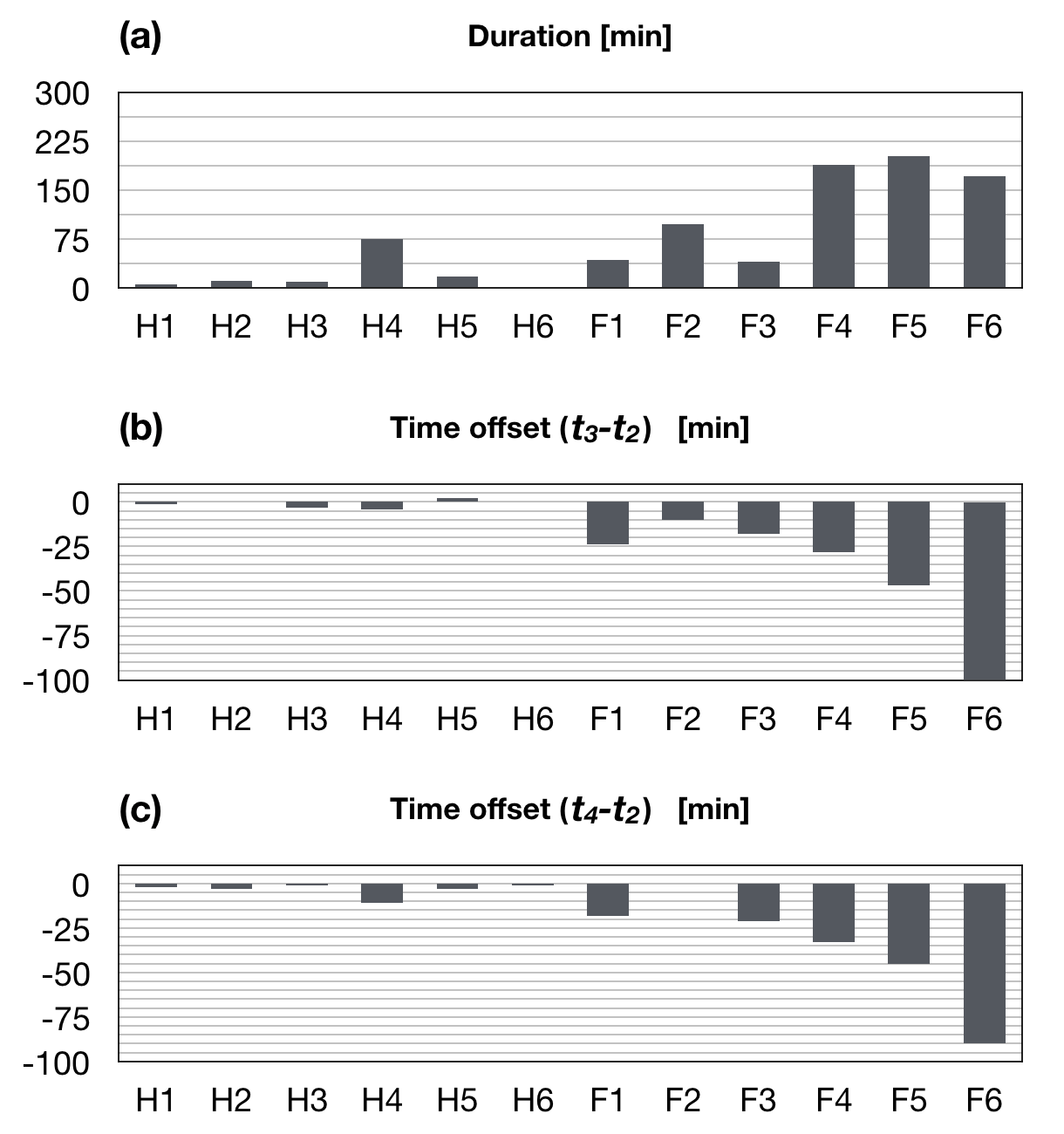}}
\caption{(a) Distribution of the duration of the slow-rise phase. (b) and (c) Distributions of the temporal offset between the onset of the main acceleration and the associated impulsive flare; negative values indicate a delayed flare.}
\label{para1}
\end{figure}

\subsubsection{Slow-rise Phase}
The velocity evolution of the hot channels shows some synchronization with the temporal variation of the integrated AIA 131~{\AA} flux also during the slow-rise phase (except H4). Both tend to show a slow rise in roughly the same time interval, with the evolution being quite close for some events (H2, H5, H6). However, two very different types of evolution are seen for the filaments. The integrated AIA 304~{\AA} flux of F2--F4 slowly increases, quite well synchronized with the slow rise of the filament velocity, similar to the hot channels. For the other three events, the integrated AIA 304~{\AA} flux decreases with time, which is due to a spreading of the erupting filament material to cover a larger area, thus resulting in more absorption. However, through carefully examining the 304~{\AA}, 171~{\AA}, and 211~{\AA} images, we find some EUV bright points or small-scale brightenings that appear near the barbs of, or underneath, all slowly ascending filaments, indicating a related occurrence of reconnection. Such brightenings can even appear for a long time prior to the beginning of the slow-rise phase. The time of the first detected brightening is shown as $t_{1}$ in Table~\ref{tb3}. The brightenings are not visible in the integrated EUV flux curves because they are small and their flux changes slowly. This indicates that the underlying reconnection evolves at small scales and in a gentle way, as expected for slow tether-cutting reconnection. 

\subsubsection{Onset Time}
Next we compare the onset times of the CME main-acceleration and impulsive flare phases. In the NOAA reports, the flare onset time is defined as the first minute in a sequence of 4~minutes of successive increase in the \textsl{GOES} 1--8~{\AA} SXR flux. Here, we obtain the onset time by carefully inspecting the temporal variation of the 1--8~{\AA} SXR flux, using the criterion that, from this time onward, the flux increases continuously and far more rapidly than before. For H1 and H6, the 15--25 keV hard X-ray flux from \textit{RHESSI} is also utilized, using the same criterion \citep[for H1 see Figure 10 in][]{cheng13_driver}. The onset of the flares associated with the quiescent filaments is determined using the same strategy, with the integrated AIA 304~{\AA} flux replacing the 1--8~{\AA} SXR flux. The results are shown as the vertical lines in red in Figures~\ref{ht}, \ref{ht1} and \ref{ht2} and are also listed as $t_{2}$ in Table~\ref{tb3}. The onset time of the main acceleration is approximated by the break point $t_3$ of our best fit with $t_0=0$. In addition, we also estimate the main-acceleration onset directly from the acceleration-time profiles as the time when the acceleration starts to increase continuously and with a magnitude larger than the standard deviation of the acceleration during the slow rise phase; this time is shown by the vertical lines in black in Figures~\ref{ht}, \ref{ht1} and \ref{ht2}, and listed as $t_{4}$ in Table~\ref{tb3}. The half period of the oscillations in the slow-rise phase may serve as a rough estimate of the uncertainty of the $t_4$ values, with the true onsets occurring earlier (not later) than $t_4$. The uncertainty lies in the range of 1--4~min for the hot channels ($\sim$8~min for H4) and of 5--25~min for the quiescent filaments, comparable to the uncertainties of $t_3$. Similarly, the true onset of the impulsive flare phase is masked by precursor activities in the slow-rise phase of several events and occurs before the listed values of $t_2$, with an uncertainty comparable to that of $t_4$ (or bigger in complex events like H4 and H1).

We then calculate the time difference between the main-acceleration onset and the flare onset ($\delta t_{32}=t_3-t_2$ and $\delta t_{42}=t_4-t_2$) as shown in Table~\ref{tb3} and Figures~\ref{para1}(b) and (c). It is found that the main-acceleration onset times derived from the fit ($t_3$) and from the acceleration-time profile ($t_4$) are very close to each other, except for H4. The difference to the flare onset time from both estimations is found to be close to zero (with a scatter of 3~min, which is smaller than the estimated uncertainty of 4~min for $t_3$) for the other five hot channel events. Note that the H4 event has no synchronization in the slow-rise phase and the poorest synchronization of all events in the main-acceleration phase. However, for most of the quiescent filaments, the onset of the main acceleration occurs much earlier than that of the flare. The indication of a delayed flare onset is weak ($\delta t_{32}$ comparable to the uncertainty of $t_3$) only for F2. For the other erupting filaments, the delay clearly exceeds the estimated uncertainty of $t_3$ ($\sim$10~min) and also the uncertainty of $t_4$ for each event, and even reaches $\approx$100~min for F6. These large delays indicate that the mechanism that initiates the impulsive rise of the associated flare (fast ``flare reconnection'') cannot be the mechanism that initiates the main acceleration of the erupting quiescent filaments.

\begin{figure*}
 \center {\includegraphics[width=12cm]{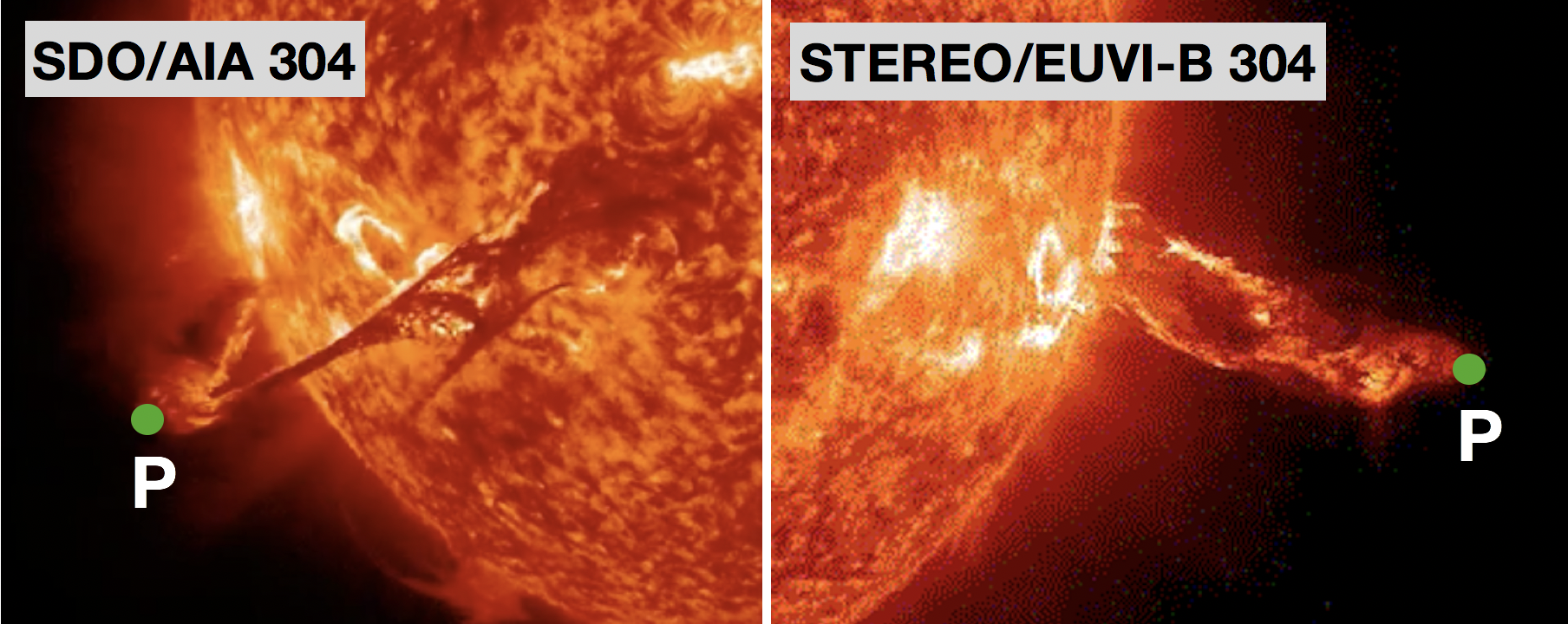}}
\caption{\textsl{SDO}/AIA 304~{\AA} and \textsl{STEREO}/EUVI 304~{\AA} images showing the erupting quiescent filament F1 as observed from two perspectives. The point P represents the same feature, identified to determine the location of the filament top in 3D at 19:45~UT on 2012 August~31.}
\label{aia_euvi}
\end{figure*}

\begin{figure*}
 \center {\includegraphics[width=14cm]{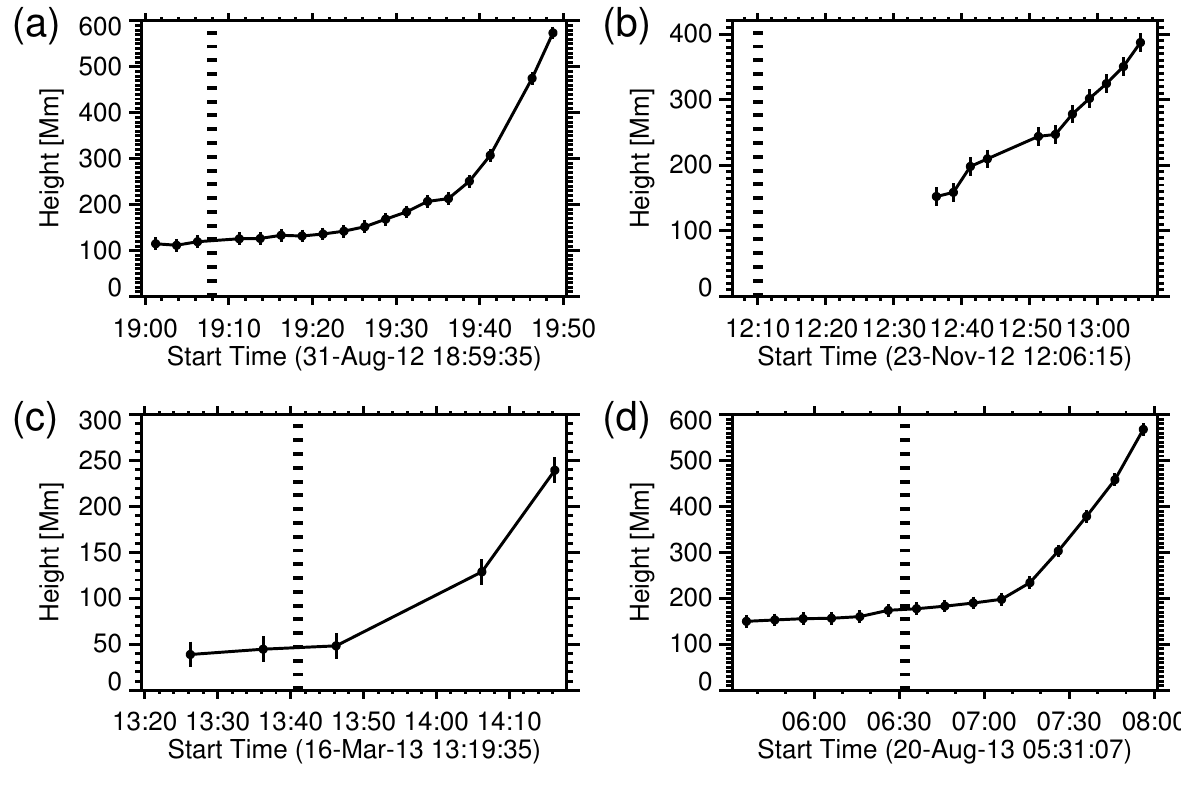}}
\caption{True (3D) height of the erupting filaments F1--F4 vs.\ time. The vertical lines denote the onset time of the main acceleration, $t_3$, with the horizontal width denoting the uncertainty.}
\label{3d_ht}
\end{figure*}

\begin{table*}
\caption{Properties of early kinematics of 12 eruption events.}
\label{tb3}{
\hspace{-0.18\textwidth}
\scalebox{0.85}[0.85]{
\begin{tabular}{cccccccccccccccc}
\\ \tableline \tableline
 Events & $t_{1}$  & $t_{2}$    & $t_{0}$ & $t_{3}$ & $t_{s}$  & $\delta t_{32}$ & $t_{4}$  & $\delta t_{42}$  & $D$ & $v$                &$a$               & $h_{c}$     & $n_{c}$ & $h_{0}$           & $n_{0}$ \\
             &  [UT]     &   [UT]      &   [UT]    &   [UT]  &   [UT]    &[min]  &   [UT]     &[min]                  &[min] & [km s$^{-1}$] &[m s$^{-2}$]  & [Mm]         &               & [Mm]              &                  \\
\tableline
H1     & 03:28  & 03:37 & 03:30 & 03:36   &03:34   &--1   & 03:35    &--2     &5    &39.5$\pm$2.9     &157.7$\pm$73.7       &50   & 1.60$\pm$0.03   &36   & 1.37$\pm$0.04   \\
H2     & 20:30  & 20:46 & 20:39 & 20:46   &20:44   &0     & 20:43    &--3     &11   &16.2$\pm$1.9     &24.6$\pm$25.8         &29   & 1.46$\pm$0.08   &19   & 1.06$\pm$0.10  \\
H3     & 09:00  & 10:28 & 10:22 & 10:25   &10:18   &--3   & 10:27    &--1     &9    &20.6$\pm$1.6     &39.7$\pm$24.3         &70   & 1.88$\pm$0.03   &60   & 1.69$\pm$0.04  \\
H4     & 00:30  & 02:05 & 00:52 & 02:01   &01:38   &--4  & 01:56    &--11   &75  &12.7$\pm$0.1      &0.4$\pm$0.2            &124    & 1.68$\pm$0.18   &36   & 1.08$\pm$0.46  \\
H5     & 12:15  & 12:30 & 12:14 & 12:32   &12:21   &2    & 12:27    &--3     &18   &45.1$\pm$1.7      &19.8$\pm$13.7        &54   & 1.57$\pm$0.10   &5     & 0.21$\pm$0.12   \\
H6     & 00:20  & 00:41 & 00:39 & 00:41   &00:40   &--0  & 00:40    &--1     &2     &22.7$\pm$12.9    &-80.6$\pm$934.3    &21   & 1.57$\pm$0.34  &18   & 1.46$\pm$0.47  \\
\tableline
F1     & 17:20  & 19:32 & 18:44 & 19:08   &18:50   &--24    & 19:14    &--18   &43     &3.1$\pm$0.3   &-0.9$\pm$1.1             & 120    &1.51$\pm$0.24   &101  & 1.34$\pm$0.27  \\
F2     & 09:40  & 12:20 & 10:38 & 12:10   &11:47   &--10    & 12:20    &--0     &97     &1.9$\pm$0.1   &0.4$\pm$0.2              & 100    &1.00$\pm$0.40   &79    & 0.89$\pm$0.35  \\
F3     & 13:00  & 13:59 & 13:09 & 13:41   &13:27   &--18    & 13:38    &--21   &40     &3.1$\pm$0.5   &--0.2$\pm$1.8            & 50     &1.05$\pm$0.56   &40    & 0.86$\pm$0.53 \\
F4     & 04:00  & 07:00 & 03:36 & 06:32   &05:41   &--28    & 06:27    &--33   &188   &2.4$\pm$0.1   &0.3$\pm$0.0              & 180    &1.25$\pm$0.23   &131  & 1.04$\pm$0.24 \\
F5     & 17:00  & 21:46 & 18:27 & 20:59   &20:30   &--47    & 21:01    &--45   &202   &2.5$\pm$0.0   &0.0$\pm$0.0              & 120   &1.42$\pm$0.16    &63    & 0.75$\pm$0.13  \\
F6     & 11:30  & 15:24 & 10:56 & 13:41   &12:15   &--103  & 13:54    &--90  &171   &2.0$\pm$0.1   &0.1$\pm$0.0               & 140    &0.92$\pm$0.11    &95    & 0.75$\pm$0.14 \\
\tableline
\end{tabular}}}\\

\vspace{0.01\textwidth}
Notes:\\
$t_{1}$ denotes the time of the first appearance of EUV brightenings, $t_{2}$ refers to the onset time of the \textsl{GOES} 1--8~{\AA} flux (integrated AIA 304~{\AA} flux) of flares associated with hot channel (quiescent filament) eruptions, $t_{0}$ is the onset time of the exponential component ($h_2(t)$) or power law component ($h_4(t)$) (see text for its relevance), $t_{3}$ is the break point between linear and nonlinear rise, approximating the onset time of the main acceleration, $t_{s}$ is the time where the exponential component velocity $v(t_{s})=v(t_{3})/10$, $\delta t_{32}$ denotes $t_{3}$--$t_{2}$. $t_{4}$ is the onset time estimated directly from the acceleration-time profile, $\delta t_{42}$ denotes $t_{4}$--$t_{2}$. $D$ and $v$ are duration and linear velocity of the slow-rise phase, $a$ is the acceleration of the quadratic fit. $h_0$ and $h_\mathrm{c}$ are initial and critical height at the onset time of the slow-rise and main-acceleration phase ($t_3$), respectively. $n_0$ and $n_\mathrm{c}$ are the corresponding decay index values of the extrapolated background field. 
\end{table*}

\subsection{Relevance of Torus Instability}\label{ss:TI}

To investigate the relevance of the torus instability in initiating solar eruptions, we compare its theoretical threshold (critical decay index of the background/strapping field) with the observationally estimated values at the onset of our events. The decay index is defined as 
\begin{equation}
\centering
n(h)=-\frac{d(\ln B_\mathrm{t})}{d(\ln h)},
\end{equation}
where $B_\mathrm{t}$ is the horizontal component of the coronal background field. For the observational estimate, the critical (i.e., onset) height and a coronal magnetic field model are required. Different methodological approaches are possible for each of them, e.g., the different fit functions used in Section~\ref{ss:kinematics} and different extrapolation schemes. Moreover, the lack of magnetic measurements from \textsl{STEREO} enforces adopting a compromise between the accuracy of the height-time and magnetic measurements. Accordingly, we have choosen different strategies for the hot channel and quiescent filament eruptions, as detailed in the following. Their respective advantages and limitations are discussed in Section~\ref{ss:discussion_decayindex}. 

First, we address the onset of the main-acceleration phase, using $t_3$ from the best height-time fit for each event as the onset time. For the hot channel eruptions, we use the corresponding height projected in the plane of sky as the critical height $h_\mathrm{c}$, but reference it to a point in the middle of the associated flare ribbons or at the bottom of the flare loops (Figures~\ref{aia}(a) and \ref{ht}(a)). Except for H4, the hot channels are sufficiently close to the limb to give a negligible difference between the projected and radial distances to the reference point. The projected height of H4 is corrected assuming a radial direction of the eruption. 

For the quiescent filaments, the critical heights are estimated through observations from two perspectives. We use the routine \texttt{scc\_measure.pro} in the SSW package, which returns the location of the filament top in 3D, including its true height, latitude, and longitude. Figure~\ref{aia_euvi} illustrates the 3D position measurement for the top section of F1. Figure~\ref{3d_ht} shows the evolution of the true heights for F1--F4. The critical heights of the F1, F3 and F4 eruptions are directly determined at the onset time, $t_3$, of the main acceleration. For the F2 and F5 eruptions, they are estimated through a backward extrapolation from the first 3D height point to the onset time, assuming a linear velocity in the early part of the main-acceleration phase not covered by \textsl{STEREO} data, and with the projection effects corrected using the true eruption direction. For F6, with only two frames of the eruption captured by \textsl{STEREO}, the projection correction is done assuming that the eruption is along the radial direction.

The relevant (external poloidal) component of the coronal background field is approximated by a potential-field extrapolation from the best available magnetogram.  We consider the Green function method to be most appropriate at the rather small onset heights of the hot channel eruptions H1--H3, H5 and H6 and the potential-field source-surface model \citep[PFSS;][]{schatten69}, which includes the influence of the heliospheric current sheet, to be most appropriate at the much larger onset heights of the quiescent filament and H4 eruptions. 
 
\begin{figure}
 \center {\includegraphics[width=8cm]{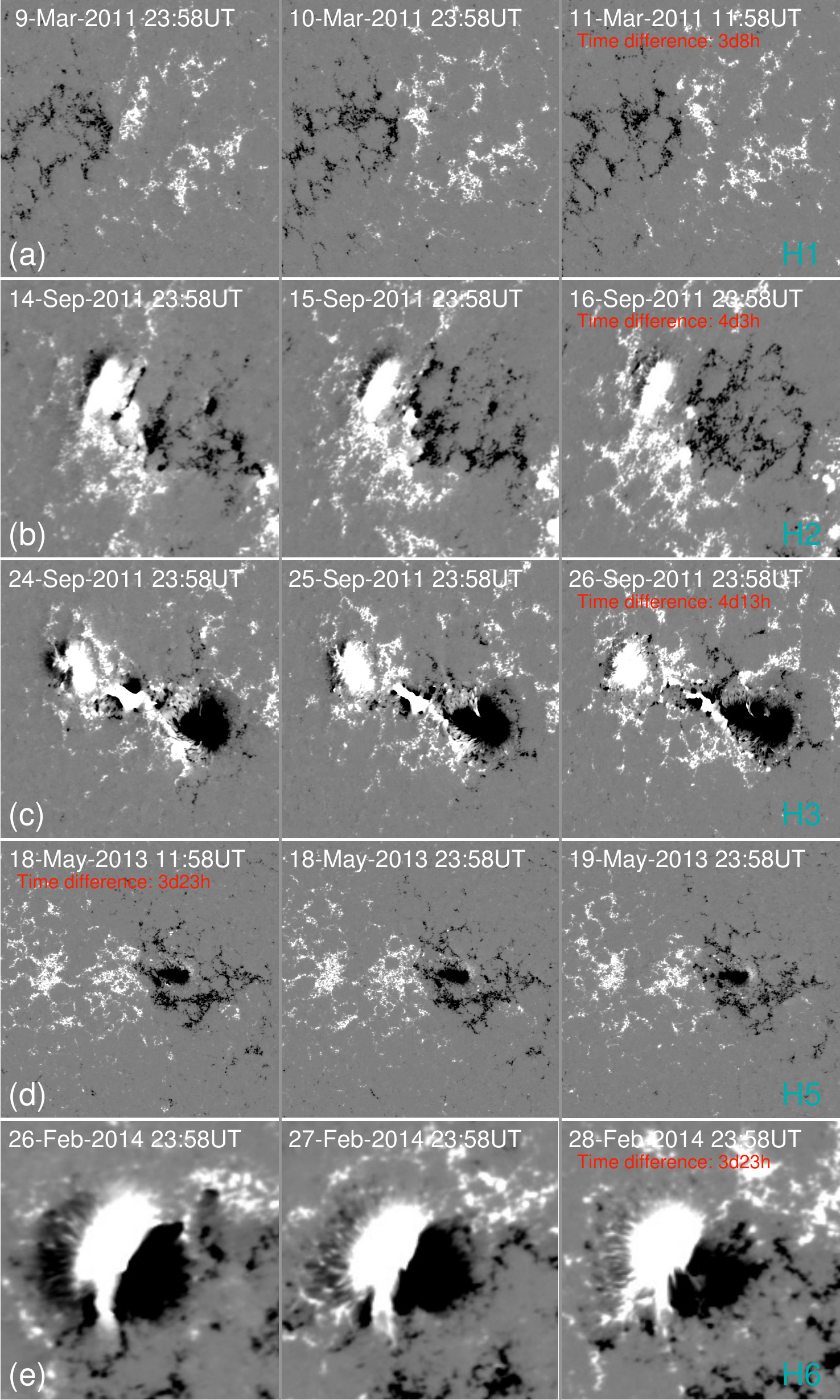}}
\caption{HMI line-of-sight magnetograms showing the evolution of source region magnetic field of H1, H2, H3, H5 and H6 during the period of 3--4 days. The magnetograms used for calculating the background field and its time intervals with the eruptions are also indicated by the time difference in red.}
\label{mag}
\end{figure}

For both groups of events, the magnetogram data necessarily contain measurements partly or fully taken at times shifted from the eruptions. This problem is more severe for our five hot channel eruptions at or near the solar limb (i.e., except H4). Their boundary data are taken 3--4 days before or after the eruptions. Our selection of events solely from active regions at least several days after their emergence phase minimizes the effect of magnetogram evolution during this period. Additionally, we argue that it is the large-scale structure of the source region which is most relevant for the determination of the decay index at the typical onset heights (see Section~\ref{ss:discussion_decayindex}), and the large-scale structure does not show strong changes during the relevant period for any of our hot channel events.  The specific evolution of the source regions can be seen in Figure~\ref{mag}. For H1 and H5, all changes in the magnetogram (emergence, shearing, dispersal, and cancellation of flux) are very minor. For H2, H3 and H6, significant flux cancellation occurs in the center of the active regions, but the main flux concentrations contributing the background field evolve only moderately, with the large-scale structure, in particular their distance, changing weakly. Therefore, we consider the inferred critical decay index values to provide a reasonable approximation of their true values. The deviations from the true decay index values are likely to increase the range found in our sample of hot channels, but not likely to introduce a false systematic trend which could strongly affect the average value. This is supported by the fact that the inferred decay index values for H1 and H5, which show the slowest magnetogram evolution, lie very close to the average value for all hot channels (see below and Table~\ref{tb3}). We base our conclusions on the average values for the two groups of eruptions.

\begin{figure}
 \center {\includegraphics[width=8cm]{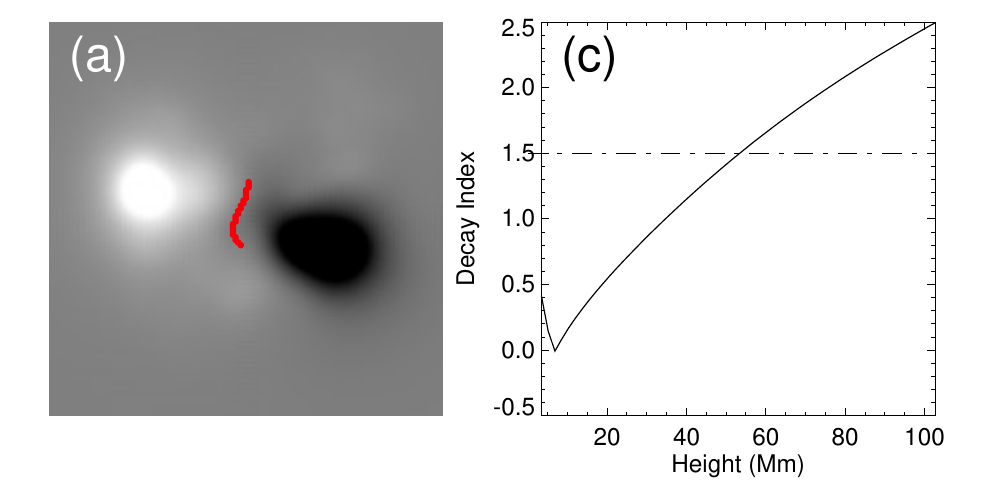}}
 \center {\includegraphics[width=8cm]{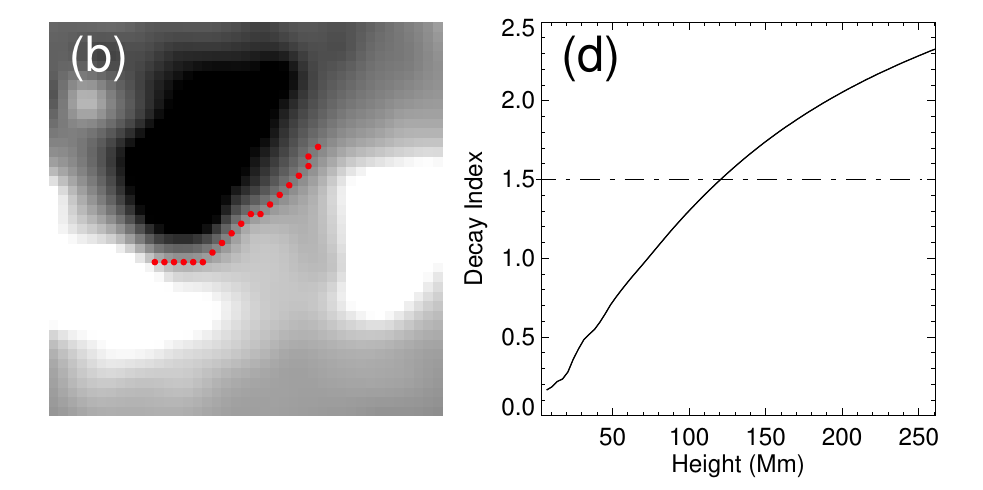}}
\caption{(a) and (b) Distribution of vertical magnetic field ($B_z$) in the H3 and F1 source regions at the height of 1.1~R$_{\odot}$. The red dotted lines show the section of the PIL included in averaging the decay index height profile, shown in (c) and (d). The theoretical value of 1.5 for the critical decay index of a toroidal flux rope is indicated by the dash-dotted lines.}
\label{index}
\end{figure}

\begin{figure*}
 \center {\includegraphics[width=15cm]{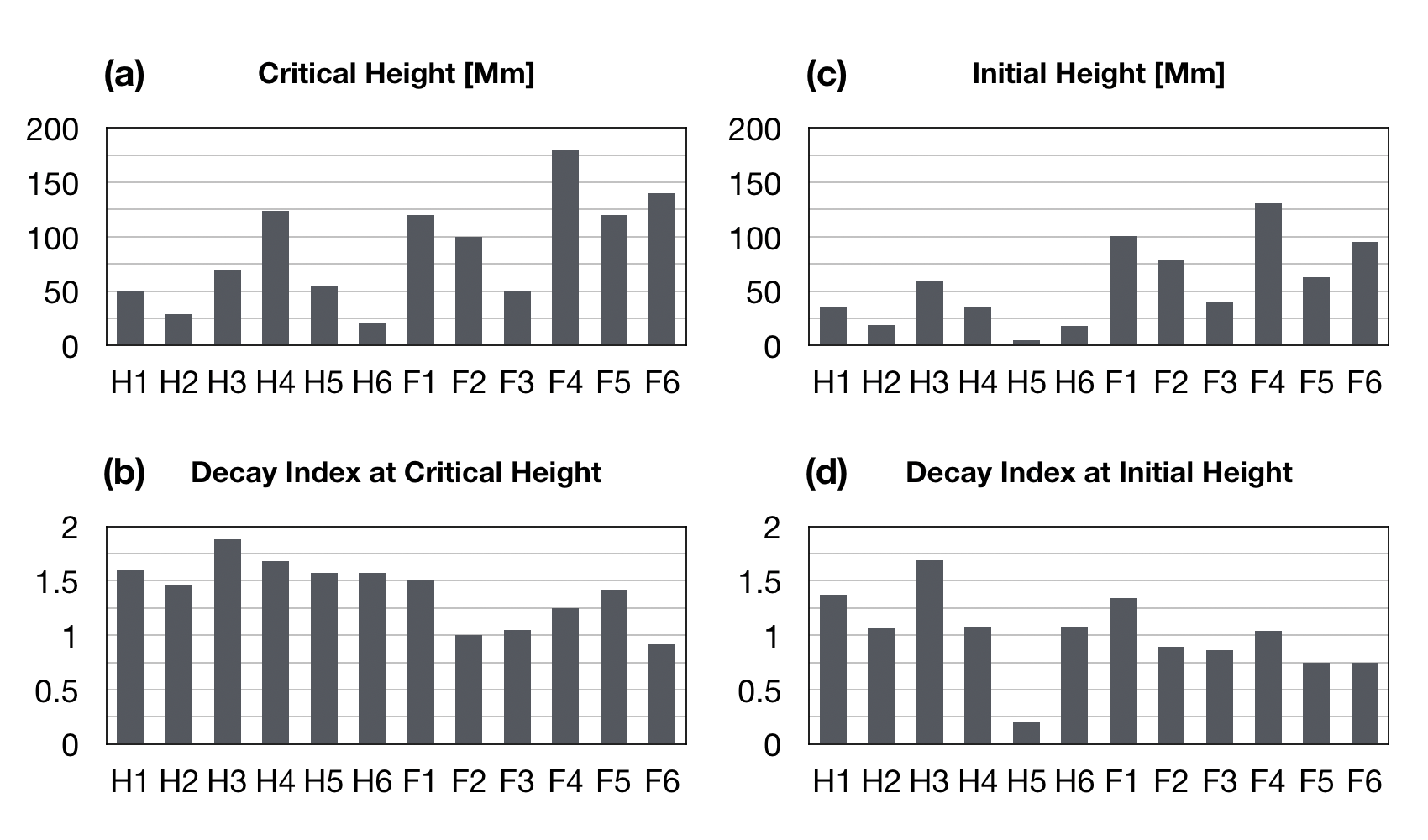}}
\caption{Distributions of (a) the critical height $h_\mathrm{c}$ and (b) the corresponding decay index $n_\mathrm{c}$ at the onset of the main-acceleration phase. (c) and (d) Initial height $h_0$ and corresponding decay index $n_0$ at the onset of the slow-rise phase.}
\label{para2}
\end{figure*}

For the quiescent filaments and H4, we use daily updated synoptic maps as the bottom boundary condition; here a 60 degree longitudinal window is updated using the average of 20 magnetograms from the same day \citep{sunxd18}. All our quiescent filament eruptions originated from within this longitude range. Therefore, the magnetogram information for part or all of the sources of the background field is updated daily. The above general arguments in favor of the meaningfulness of the derived decay indices for the hot channels, especially of their average value, apply here as well.

The results for H3 and F1 are shown in Figure~\ref{index}. Panels (a) and (b) display the distributions of the vertical magnetic field component at the height of $1.1~R_{\odot}$. We determine the relevant section of the main polarity inversion line (PIL), shown by the dotted lines, at this height in the middle of our eruption onset heights, rather than at the photospheric level. Panels (c) and (d) show the decay index vs.\ height, averaged along the relevant section of the PIL. The inferred critical heights ($h_\mathrm{c}$) and corresponding decay index values ($n_\mathrm{c}$) for all events are compiled in Table~\ref{tb3} and Figures~\ref{para2}(a) and (b). The decay index errors are the standard deviations of all decay index values above the selected pixels along the PIL. 

From Table~\ref{tb3} and Figure~\ref{para2}(a), one can see that the critical heights for the hot channel eruptions are distributed in the range of 21--83~Mm with an average of 50~Mm, which are systematically smaller than those for the quiescent filament eruptions, 50--180~Mm with an average of 118~Mm. This corresponds to the different spatial scales of the main flux concentrations in the photospheric boundary. However, the decay indices at the critical heights for the former, ranging from $1.46\pm0.08$ to $1.88\pm0.03$ with an average $\overline{n_\mathrm{c}}$ of 1.6$\pm0.1$, are systematically greater than those for the latter, which fall in the range of $0.92\pm0.11$--$1.51\pm0.24$ with an average $\overline{n_\mathrm{c}}$ of 1.2$\pm0.2$ (Figure~\ref{para2}(b)). These decay indices for the hot channels are close to the threshold of the torus instability for the circular flux rope \citep[1.4--1.9;][]{torok05,kliem06,fan07,aulanier10}, and the values for the quiescent filaments are comparable with the threshold of the torus instability for the straight flux rope \citep[1.1--1.3;][]{demoulin10}.

The initial heights ($h_0$) and resulting decay index values ($n_0$) at the onset of the slow-rise phase tend to be subcritical, as can be seen in Table~\ref{tb3} and Figures~\ref{para2}(c) and (d). For the hot channels, the initial height represents the height where they can first be identified obviously. For the filaments, the initial height is deprojected using the expression $h_0=h_\mathrm{c} h_\mathrm{p0}/h_\mathrm{pc}$, where $h_\mathrm{p0}$ and $h_\mathrm{pc}$ are the projected heights at the first and break point, respectively. We find that the initial heights yield clearly lower decay index values in the ranges $n_0=0.21\pm0.12\mbox{--}1.69\pm0.04$ ($\overline{n_0}=1.1\pm0.5$) for the hot channels and $n_0=0.75\pm0.14\mbox{--}1.34\pm0.27$ ($\overline{n_0}=0.9\pm0.2$) for the quiescent filaments. The averages clearly fall below the instability threshold. It is worth noting that these are upper limits for both the onset heights and corresponding decay indices, because the slow-rise phase might actually commence before the first measured data point \citep[e.g.,][]{xing18}.

The inferred decay index values, in particular their averages, suggest that the main-acceleration phase commences by the onset of the torus instability. This also naturally explains the initially exponential evolution of the acceleration in this phase for the majority of the events and is consistent with the power-law evolution for the remaining events \citep{schrijver08_filament}. In comparison, the onset of the slow-rise phase by ideal MHD instability, as considered in \citet{zhang06}, see their Figure~1, is much less supported because the corresponding decay index values mostly lie below 1.1, which is the smallest threshold derived so far for the torus instability \citep{demoulin10}. 

\section{Discussion}\label{s:discussion}
\subsection{Early Kinematics}\label{ss:discussion_kinematics}
In this paper, we study the initiation and early kinematic evolution of 12 solar eruptions including 6 active region hot channel eruptions and 6 quiescent filament eruptions. The 12 events produce 11 CMEs with a wide velocity distribution, ranging from $\sim$500 km s$^{-1}$ (for F2 and F6) to $\sim$2000 km s$^{-1}$ (for H3 and H6); their height-time profiles, though, do not differ qualitatively. This indicates that the basic two-phase initial kinematic evolution of CMEs may be uniform in character for most events, largely irrespective of the details of the pre-eruptive configuration. 

All eruptions studied here exhibit a slow-rise phase followed by a main-acceleration phase, similar to the previous results derived from EIT and LASCO data at low cadence \citep[e.g.,][]{zhang01,neupert01,sterling07}. In the slow-rise phase, the erupting structures rise with an approximately linear behaviour. The acceleration is extremely small or not even measurable for nine of our 12 events, small for two events (H3 and H5), and moderate ($\sim\!160\pm70$~m~s$^{-2}$) only for H1. The main-acceleration phase starts with a rapidly and nonlinearly increasing acceleration, indicating instability. Through experimenting with linear, quadratic, exponential, and power-law functions, we find that the ones consisting of a nonlinearly accelerating component superimposed with a linear component fit the height-time profiles best for all events. From the superiority of the superposed fit functions, as well as from the shapes of the acceleration-time profiles, we conclude, opposite to \citet{kahler88}, that the main-acceleration phase is qualitatively different from the slow-rise phase, strongly suggesting that different physical mechanisms govern them. This implies that ``runaway reconnection", conjectured in the original tether-cutting model \citep{moore01}, can not be a uniform mechanism for both phases. However, this model remains a possible model for the slow-rise phase, as suggested by many recent investigations, including the present work. 

The nonlinear rise is approximately exponential for three of our six hot channel eruptions and for all six filament eruptions, while a power law describes the rise best for the remaining three hot channel eruptions. This is consistent with the result of \citet{vrsnak01}, in which an exponential-like or power-law-like growth of the CME height is also found for a large sample of events, even valid in the higher corona. By contrast, \citet{schrijver08_filament}, considering two eruptions from active regions, concluded that the power law function can yield a slightly better fit than the exponential in the main-acceleration phase. From our fit results and from the relevance of the torus instability (Section~\ref{ss:TI}), we conjecture that the nonlinear rise is mostly exponential for the majority of eruptions. Considering the different conclusion in \citet{schrijver08_filament}, there appear to exist three possible reasons for the superiority of a power-law fit. First, the rise can indeed be closer to a power law when an instability is triggered by a sizeable perturbation (e.g., a sympathetic eruption or rapid flux emergence), as demonstrated in \citet{schrijver08_filament}. However, such a scenario does not appear to be very frequent; rather, the photospheric evolution toward an eruption is usually gradual, and sympathetic events are rare. Second, the fit functions employed in \citet{schrijver08_filament} gave an advantage to the power law, which contained one more free parameter than the exponential, thus allowing for a higher flexibility. Third, and probably most important, a power law should be preferred if the fitting includes some data points beyond the linear phase of an instability into the saturation phase (characterized by a decreasing acceleration), as done in \citet{schrijver08_filament}. Since the exponential rises faster than the power law in the long run, it should yield a stronger deviation from the data points in the saturation phase. The number of data points in the main-acceleration phase is limited by its duration relative to the cadence of the observation. Not surprisingly, the quiescent filament eruptions, which provide many data points before the saturation sets in, all favor the exponential fit. Further investigation of the kinematics of eruptions from active regions is required to clearly elucidate the relevance of exponential vs. power-law behaviour.

\subsection{CME-flare Timing Relationship}\label{ss:discussion_timing}
We also study the timing relation of the evolution of CMEs to that of the associated flares. With only one exception (H4), the velocity in the main part of the CME main-acceleration phase is synchronized with the SXR and EUV light curve in the main part of the flare impulsive-rise phase in our sample. The synchronization is rather close for the majority of the events (H2, H3, H5, H6, F1, F4--F6), even very close for some of these, and moderate for the others (H1, F2, F3). In the H4 event, the hot channel evolves with a much larger volume than the flare, so the synchronization is poor, but a flare associated in time with the hot channel eruption exists as well. The timing of the peak CME acceleration relative to the peak flare energy release rate can serve as a quantitative measure of the synchronization between CME acceleration and flare energy release \citep{maricic07}. However, for many of our events, we cannot reliably determine the point of peak acceleration, because the hot channels tend to fade, and the quiescent filaments tend to approach the edge of the AIA field of view, before the peak acceleration is reached.

A synchronization, albeit less close, also exists in the slow-rise phase of the hot channel eruptions (again, except for H4), which then already show weak flare signatures. The integrated AIA 304~{\AA} emission of the source region, taken as a proxy for a flare signature in the slow-rise phase of our quiescent filament eruptions, shows a synchronization with the slow-rise velocity in half of the events. The opposite trend is noticed in the other half; here the enhanced absorption by the spreading of the filament masks any synchronization that might exist. 

The onset times of the CME main acceleration and flare impulsive rise are found to be very close to each other for five hot channel eruptions (except the complex event H4). On the other hand, five of the six quiescent filament eruptions show a delayed onset of the impulsive flare phase. The delay is unambiguous for F4, F5, and F6, very likely for F1 and F3, and weakly indicated for F2. For F1 and F6, it is unlikely that the background level of the 304~{\AA} emission from the source region is so low that its flare-related rise would start as early as the time $t_3$ derived for the CME onset (see Figures~\ref{ht}(d) and \ref{ht2}). The delays for these five events are substantial, lying in the range of 18--103~minutes. These results are consistent with those in \citet{maricic07}, who found a significantly delayed flare onset (by $\ge30~\%$ of the duration of the main acceleration) for 6 out of 18 eruptions, but mostly from active regions. 

Overall, these findings agree very well with previous results that the main CME acceleration and main flare energy release, the latter being due to reconnection, are coupled \citep[e.g.,][]{zhang01}. They also indicate an association between the preceding slow ascent of flux and reconnection in the source region, except in those cases where a filament spreading masks any potentially related increase of the EUV flux (F1, F5, and F6). This is consistent with the conjecture that tether-cutting reconnection is an important process in this phase. In spite of these associations, the fitting also yields a clearly delayed onset of the flare impulsive phase for the majority of our quiescent filament eruptions. The latter result, although requiring substantiation from a larger sample of events, favor instability models above reconnection models for the \emph{onset} of the main CME acceleration. This is true, in particular, if the same mechanism initiates the eruptions of quiescent filaments and from active regions, as is widely assumed. 

\subsection{Relevance of Torus Instability for Eruption Onset and Driving}\label{ss:discussion_TIrelevance}
Based on the onset time of the main acceleration, the critical height and resulting decay index are inferred. (The reliability of these values requires discussion which we provide in Section~\ref{ss:discussion_decayindex} in the comparison with other recent inferences in the literature.) For the hot channels, the decay indices are found to be close to the threshold of torus instability for the circular flux rope \citep[1.4--1.9;][]{torok05,kliem06,fan07,aulanier10}, while for the quiescent filaments, the values cluster around the threshold of torus instability for the straight flux rope \citep[1.1--1.3;][]{demoulin10}. This is consistent with the observations that the hot channels usually present a curved loop-like structure \citep[e.g.,][]{zhang12,cheng13_driver,patsourakos13,lileping13,liting13_homologous,tripathi13,vemareddy14,chintzoglou15,joshi15,zhougp16}, whereas the quiescent filaments typically appear as a set of long and nearly straight threads, overall much closer to an only weakly bent cylinder \citep[e.g.,][]{yangshuhong14,yanxl15_quiescent_filament,liqin17}. 

These results suggest that the torus instability of a flux rope initiates the main-acceleration phase. No association of eruption onset with the critical decay index of the torus instability is expected if magnetic reconnection is the initiating process. The onset of ideal MHD instability also naturally leads to an 
exponential growth of the eruptions and to the development of a feedback between the instability and reconnection during the main-acceleration phase. Once the instability sets in, the flux rope is driven to erupt outward rapidly. With a short time delay, the fast flare reconnection, taking place in a narrow, long stretching current sheet formed below the erupting flux rope \citep{linjun05,savage10,liuw13,ling14,sun15_nc,cheng18_cs}, is switched on or strongly amplified from preceding slow tether-cutting reconnection. This produces two relevant effects: an enhancement of the force imbalance due to the transfer of the overlying flux into poloidal flux of the flux rope and the upward slingshot effect of the reconnected flux. Both of them provide an upward force that can additionally accelerate the erupting CME, which, in turn, further facilitates the development of the torus instability. That is to say, the main-acceleration phase is \emph{expected} to be a process consisting of a combination of ideal torus instability and magnetic reconnection in a positive, mutually amplifying feedback. 

Unfortunately, at present, it is still extremely difficult to figure out which mechanism (torus instability or magnetic reconnection) provides a dominant contribution in the main-acceleration phase. This might even vary from event to event. The hot channels tend to lose their equilibrium at relatively low heights. Fast reconnection then tends to commence promptly. As the involved field is strong, the two reconnection-induced effects should be very efficient, especially if the reconnection sets in almost simultaneously in a relatively elongated area below the rising flux rope. In such cases, the sudden transfer of the overlying flux to poloidal flux of the rope immediately amplifies the acceleration strongly. Similarly, the sudden upward snapping of reconnected field lines may also efficiently contribute to the acceleration of the eruption. 

For the quiescent filament eruptions, very weak flares are typically associated, manifesting as ribbons and post-flare arcades only in the 304~{\AA} and 171~{\AA} passbands. This is also common for polar crown prominence eruptions \citep{song13, gopal15}. The corresponding CMEs tend to be slower, because they originate in larger source regions with weaker magnetic field. The main acceleration starting earlier than the flare onset suggests that the acceleration process may be first dominated by the torus instability, as the reconnection and its induced two effects should not be very efficient (weak fields and relatively large heights at which reconnection occurs). The slingshot effect, for example, turns in such cases into a weak reconnection outflow, which may deform the flux rope a little bit if it catches up to the erupting rope, but may not accelerate it considerably. Similarly, the force imbalance due to flux transfer should not be very strong as well. The distance between the flux rope and the reconnecting X-line is larger and the reconnection jet velocity is smaller compared to eruptions from active regions. Therefore, if the flare reconnection is a consequence of the ideal MHD instability, then one can expect that any delay of the flare onset, as well as the time needed to fully establish the feedback between ideal instability and reconnection, tend to be longer for erupting quiescent filaments, as found in the events studied here. The same trend is also indicated, albeit weakly, within our group of quiescent filament eruptions: the eruptions F4--F6 show the longest delay and two of them (F4 and F6) are the slowest eruptions in the sample when only the AIA field of view is considered (see Figure~\ref{ht2}, second column). Since any delay of reconnection onset can also depend on the magnetic topology of the source region, a close correlation with the velocity of the eruption is not expected. 

For the slow-rise phase, the torus instability is unlikely to be the universal onset and driving process because the decay index at the inferred onset height is clearly sub-critical for the majority of the investigated events. We argue that the slow rise of all 12 events is closely associated with slow tether-cutting magnetic reconnection. First, this is indicated by the synchronization of the velocity-time profiles and flare light curves, which is seen whenever enhanced absorption by a filament is not dominant in the light curve. Second, small-scale EUV brightenings are seen in the source regions of all events. This supports the occurrence of reconnection in the slow-rise phase but with a much slower magnetic dissipation rate than that in the main-acceleration phase. Such a slow reconnection is very critical to create more poloidal flux and lift the hot channels gently. This is also inferred by recent work by \citet{liutie19}, in which a hyperbolic flux tube (HFT) was identified underneath the hot channels prior to their eruption. The tether-cutting type reconnection in the HFT then proceeds slowly, due to the slow driving from the photosphere, as long as no fast driving from an MHD instability in the corona occurs, simultaneously leading to the slow rise and heating of the hot channels. The slow rise of the quiescent filaments can additionally be driven by mass draining (see \citealt{Jenkins&al2019} and references therein). Such a feature is observed in the slow-rise phase of F2 but not in the other filament events studied here.

\subsection{Estimate of Decay Index}\label{ss:discussion_decayindex}
Finally, we discuss the reliablity of our estimates of the onset heights, $h_\mathrm{c}$, and critical decay index values, $n_\mathrm{c}$. It is clear that an accurate estimate of the onset height is as important as an accurate coronal field model for obtaining a reliable decay index. However, as pointed out in Section~\textbf{\ref{ss:TI}}, with current instrumentation, a compromise between the reliability of the height-time and magnetogram data must be chosen, especially for hot channels and other rapidly evolving (fast) eruptions. For events near the limb, the $h(t)$ data from \textsl{SDO}/AIA have a high accuracy and cadence, yielding the most reliable onset heights, but the magnetic data have a temporal offset of 3--4 days. For events near the disk center, the magnetic data are optimized, but \textsl{STEREO}/EUVI does not provide $h(t)$ data for hot channels and yields lower cadence for filaments. At present, it is not clear whether one of these choices or a compromise in the middle yields the most reliable $n_\mathrm{c}$ values. The former choice may be best for events from slowly evolving source regions, and the latter choice may be best for very slowly rising eruptions.

Our sample of quiescent filaments erupting from longitudes $\le\!60^\circ$ (Table~\ref{tb1}) represents a good compromise for this category of events. This choice yields accurate and reliable onset heights from the combination of AIA and EUVI data and mostly very reliable and nearly up-to-date magnetograms of the source region. On the other hand, the temporal offset of the magnetogram data for the five hot channel eruptions near the limb introduces an uncertainty of the computed coronal field and inferred decay index values. We argue, however, that substantial errors in the decay index values are unlikely, because the decay index at the relevant onset heights is primarily determined by the large-scale structure of the active-region magnetograms. The typical height for torus instability onset is about the half-distance, $L_\mathrm{f}$, between the main photospheric flux concentrations that provide the background field (this is where $n=1.5$ in a bipole). At this height, the large-scale structure of the photospheric field at scales $\sim L_\mathrm{f}$ determines the structure of the coronal background field and its decay index. Typically, the large-scale structure changes only slowly for active regions after their emergence phase, i.e., at most moderately in the given time span. Figure~\ref{mag}, discussed in Section~\ref{ss:TI}, confirms that the large-scale structure does not show strong changes during the relevant period for any of our hot channel events from the limb. The flux cancellation seen in the evolution toward the eruptions H2, H3, and H6 at a scale $\ll L_\mathrm{f}$ will influence $n(h)$ primarily at scales $h\ll L_\mathrm{f}$ and only weakly at $h\sim h_\mathrm{c}$. Although the measured field strengths become less reliable closer to the limb, the geometric evolution can still be judged, especially the evolution of $L_\mathrm{f}$. The effect of magnetogram evolution on the decay index values in the relevant height range ($h\sim L_\mathrm{f}$) can be quantitatively studied using a sample of eruptions from active regions near central meridian. This will be done in a follow-up study, to support the methodology chosen here. 

To illustrate the complexity of the methodological approach to the problem, including the very important role of precise $h(t)$ data, it is instructive to compare our values with those published very recently by \citeauthor{Vasantharaju2019} (\citeyear{Vasantharaju2019}, henceforth V19), \citeauthor{zou19} (\citeyear{zou19}, henceforth Z19), and \citet{Myshyakov20}. V19 and Z19 determined the onset of the main-acceleration phase of erupting filaments/prominences using the same fit function $h_2(t)$ with $t_0=0$ and the same expression for the onset time (our $t_3$), and also inferred the critical decay index at the obtained onset height. V19 selected seven filament eruptions from active regions and three from between active regions, so-called intermediate filament eruptions. There is a salient difference in the results for the critical decay index, found to lie in the range 0.8--1.2, averaging to $\sim$1.0, for the active-region filaments (0.8--1.3 if the intermediate filaments are included in V19,  while our critical decay indices for the eruptions from active regions fall in the range 1.5--1.9, with an average of 1.6. On the other hand, Z19 found the critical decay index for filament eruptions from active regions to lie in the range 0.4--2.5, which is consistent with our range (although far broader), with an average of $\sim$1.5, close to our average. 

The analysis of the kinematic evolution in V19 and Z19 is very similar to ours, with probably less precise height data in V19 and Z19 because their source regions were chosen mostly on disk (in a range of intermediate longitudes $\pm40\mbox{--}80^\circ$ in V19 and without any longitude selection in Z19). This is a likely reason for the large scatter of the $n_\mathrm{c}$ values in Z19, but can hardly explain the strong systematic difference to V19. The main reason causing the latter, we speculate, is that V19 underestimated the critical height. They found the critical heights for their seven eruptions from active regions to lie in the range $h_\mathrm{c}=10\mbox{--}38$~Mm, with an average of 22~Mm. This is much smaller than our $h_\mathrm{c}=21\mbox{--}124$~Mm, with an average of 57~Mm. The main reason appears to be the fact that they included a much earlier part of the slow rise in their analysis of the kinematics, possibly due to the low cadence available from \textsl{STEREO}. The early part typically presents a very small velocity \citep[$<$1 km s$^{-1}$;][]{xing18}. Such a small velocity results in the crossing of the linear and exponential components in $h_2(t)$, i.e., the onset time, shifting toward an earlier time, thus giving rise to a smaller onset height. Additional reasons could be (1) that filament heights, as used by V19, may fall systematically below the heights of flux ropes (hot channels used here), as suggested by \citet{zuccarello16}, especially during the main-acceleration phase, (2) that several of the events in V19 actually originated in areas of rather dispersed field, (3) that most of their events originated from longitudes $>\!50^\circ$, where the daily update of the synoptic magnetogram has only a limited or no effect, (4) that the different choice of the coronal field model has a systematic effect on the decay index estimates, (5) that power-law fits may tend to yield larger onset heights, as is the case for our events H5 and H6, and (6) that the samples of active-region events still have a relatively small size. 

Specifically, V19 computed the potential field in the PFSS approximation, while we use the Green function in a Cartesian box. The latter assumes that all sources of the coronal field are localized under the magnetogram area, so that the field asymptotically decreases like a dipole field, i.e., $n\to3$ for $h\to\infty$. The PFSS approximation implies the presence of the heliospheric current sheet outside the source surface. This additional source changes the asymptotic behavior of the field to a significantly slower decrease, $n<3$. From our experience with applying the PFSS method to $>$\,50 cases, the decay index in the height range approaching the source surface scatters strongly, with values around 2.2 being most common. Irrespective of this specific value, it is clear that the PFSS model tends to drop the decay index to values lower than the ones in the corresponding potential field computed with the Green function. This is consistent also (1) with the slightly smaller average critical decay index in Z19 compared to ours and (2) with the high critical decay index values of 1.5--1.8 for three quiescent filament eruptions found by \citet{Myshyakov20}, who used the Green function to compute the coronal field. The critical values $n_\mathrm{c}$ inferred in the literature for eruptions from the quiet Sun are typically smaller than those for eruptions from active regions, similar to our results in Section~\ref{ss:TI}. While this is plausible from the geometrical difference (typically flatter vs.\ typically more arched erupting structures, respectively), an influence of the universal use of the PFSS model for eruptions from the quiet Sun, except in the study by \citet{Myshyakov20}, cannot be excluded. 

From the above discussion, it is clear that a reliable determination of the decay index at the onset of eruptions requires precise height-time measurements at high cadence, a reliable magnetogram, and an appropriate choice of the extrapolation method. All of the factors listed above, which definitely or potentially influence the inferred decay index values, will be addressed in subsequent investigations.

\section{Conclusions}\label{s:conclusions}
We obtain the following conclusions, valid in the whole range of rise velocities reached by the 12 studied eruptions. 

\begin{enumerate}
\item The fitting confirms that the slow-rise and main-acceleration phases of solar eruptions are qualitatively different, indicating different dominant mechanisms. 

\item The slow-rise phase is well approximated by a linear or quadratic ascent with the majority of events showing a very small acceleration. An obvious quadratic contribution is found only in three of our six hot channel events.

\item The main-acceleration phase is characterized by an exponential rise in the majority of events, indicating instability. For a small fraction of the events (three in our sample), the rise is closer to a power law. Further studies of data with higher cadence are required to clarify the relevance of power-law behavior in this phase. 

\item The kinematic evolution of the eruptions tends to be synchronized with reconnection in the source volume as represented by the SXR or EUV light curve of the associated flare. The synchronization is found in both the slow-rise and main-acceleration phases and is often but not always close. This indicates a strong role for slow tether-cutting reconnection in the slow-rise phase and a positive feedback between ideal MHD instability and fast flare reconnection in the main-acceleration phase. 

\item The onset times of CME main acceleration and flare impulsive rise lie close to each other for the hot channel eruptions, except for one complex event (H4). The delays scatter within 3~min, less than their estimated uncertainty. On the other hand, a delayed onset of the impulsive flare phase is found in the majority of our quiescent filament eruptions (5 out of 6) and weakly indicated in the remaining one. This delay and its trend to be bigger for slower eruptions are consistent with the conjecture that an ideal MHD instability initiates and initially drives solar eruptions. 

\item The decay index of the ambient field at the starting height of the main-acceleration phase lies close to the threshold of the torus instability for all 12 events (1.6$\pm$0.1 on average for the typically arched hot channel eruptions from active regions and 1.2$\pm$0.2 on average for the much flatter erupting quiescent filaments). This suggests that the torus instability initiates and initially drives the main-acceleration phase in the majority of solar eruptions. However, the accuracy of the decay-index calculation is limited by the lack of reliable magnetograms for five of our hot channel events that occur close to or above the limb. We believe that this limitation does not change our main conclusion, but further studies need to be pursued to confirm it. 

\end{enumerate}

Code availability: The codes used for measuring and fitting height-time data are available at the website http://spaceweather.gmu.edu/public/xcheng/fit/.

\acknowledgements  We thank the ISSI team on ``Decoding the Pre-Eruptive Magnetic Configuration of Coronal Mass Ejections", led by S. Patsourakos \& A. Vourlidas, for stimulating the research presented here. 
We also acknowledge constructive comments by the referee, which helped improving the clarity of the manuscript. 
\textsl{SDO} is a mission of NASA's Living With a Star Program. \textsl{STEREO}/SECCHI data are provided by a consortium of NRL (US), LMSAL (US), NASA/GSFC (US), RAL (UK), UBHAM (UK), MPS (Germany), CSL (Belgium), IOTA (France), and IAS (France). X.C., C.X., Z.J.Z and M.D.D. are supported by NSFC grants 11722325, 11733003, 11790303, 11790300, Jiangsu NSF grant BK20170011, the ``Dengfeng B" program of Nanjing University, and the Alexander von Humboldt foundation. B.K. acknowledges support by the DFG and NSFC through the collaborative grant KL 817/8-1/NSFC and support by NASA through grants NNX16AH87G, 80NSSC17K0016, 80NSSC19K0082, and 80NSSC19K0860. T.T. was supported by NASA's HSR program (award no. 80NSSCK0858), NSF's Solar Terrestrial program (award no. AGS-1923377), and NSF's PREEVENTS program (award no. ICER-1854790). J. Z. is supported by NASA grant NNH17ZDA001N-HSWO2R.

\end{document}